\documentclass[a4paper, 11pt]{article}
\pdfoutput=1 
 
\usepackage{jheppub} 
                     
\usepackage[T1]{fontenc} 
                     
\usepackage{hyperref}  
\usepackage{amsmath,physics,float}  
\usepackage{amssymb}  
\usepackage{latexsym}
\usepackage{graphicx,xcolor}
\usepackage{subfig} 
\usepackage{empheq}
\definecolor{pink}{HTML}{FDDBC7}
\definecolor{red}{HTML}{B2182B} 
\definecolor{blue}{HTML}{4393C3}

\usepackage{skak}

\usepackage{braket}
\usepackage{amssymb}
\usepackage{amsmath}
\usepackage{amsthm}
\usepackage{mathrsfs}
\usepackage{framed}
\usepackage{hyperref}
\usepackage{slashed}
\usepackage{bbm}

\hypersetup{colorlinks=true}
\hypersetup{linkcolor=red}
\hypersetup{citecolor=violet}
\hypersetup{urlcolor=red}
\numberwithin{equation}{section}

\usepackage{tikz}
\usetikzlibrary{arrows,shapes}
\usetikzlibrary{trees}
\usetikzlibrary{matrix,arrows} 				
\usetikzlibrary{positioning}				
\usetikzlibrary{calc,through}				
\usetikzlibrary{decorations.pathreplacing}  
\usepackage{pgffor}							

\usetikzlibrary{decorations.pathmorphing}	
\usetikzlibrary{decorations.markings}
\tikzset{
    vector/.style={decorate, decoration={snake}, draw},
	provector/.style={decorate, decoration={snake,amplitude=2.5pt}, draw},
	antivector/.style={decorate, decoration={snake,amplitude=-2.5pt}, draw},
   	 fermion/.style={draw=cyan, postaction={decorate},
        decoration={markings,mark=at position .55 with {\arrow[draw=black]{>}}}},
    fermionbar/.style={draw=cyan, postaction={decorate},
        decoration={markings,mark=at position .55 with {\arrow[draw=black]{<}}}},
    fermionnoarrow/.style={draw=black},
    gluon/.style={decorate, draw=black,
        decoration={coil,amplitude=4pt, segment length=5pt}},
    scalar/.style={dashed,draw=black, postaction={decorate},
        decoration={markings,mark=at position .55 with {\arrow[draw=black]{>}}}},
    scalarbar/.style={dashed,draw=black, postaction={decorate},
        decoration={markings,mark=at position .55 with {\arrow[draw=black]{<}}}},
    realscalar/.style={draw=black}, 
    electron/.style={draw=black, postaction={decorate},
        decoration={markings,mark=at position .55 with {\arrow[draw=black]{>}}}},
	bigvector/.style={decorate, decoration={snake,amplitude=4pt}, draw},
    phir/.style={draw=blue, postaction={decorate},},
   phil/.style={dashed,draw=blue,},
     phiav/.style={draw=cyan, postaction={decorate},},
   phidif/.style={dashed,draw=cyan,},  
    chir/.style={draw=red, postaction={decorate},},
   chil/.style={dashed,draw=red,},
   psir/.style={draw=violet, postaction={decorate},},
   psil/.style={dashed,draw=violet,},   
   	bigvectorr/.style={decorate, decoration={snake,amplitude=4pt}, draw=blue},
   	bigvectorl/.style={decorate, decoration={snake,amplitude=4pt}, dashed, draw=red},
   	vectorr/.style={decorate, decoration={snake}, draw=blue},
    vectorl/.style={decorate, decoration={snake}, dashed,  draw=red},  
}

\tikzstyle{block} = [draw, rectangle, 
    minimum height=3em, minimum width=6em]

\def\phir{\phi_{\textrm{R}}} 
\def\phil{\phi_{\textrm{L}}}

\def\im{\textrm{Im}} 
\def\re{\textrm{Re}} 
\def\ln{\textrm{ln}}

\def\msbar{ {\bar {\textrm{MS}}} }

\def\msbar{ {\overline {\textrm{MS}}} }

\newcommand{\chibarpropagator}[7]{
\draw [#6, ultra thick] (#3,#4) -- (#1/2+#3/2,#2/2+#4/2);
\draw [#7,ultra thick] ((#1,#2) -- (#1/2+#3/2,#2/2+#4/2);	
\node at (#1/2+#3/2+.5,#2/2+#4/2+.5) {#5};	
} 

\newcommand{\chibarpropagatorr}[5]{
\chibarpropagator{#1}{#2}{#3}{#4}{#5}{chir}{chir}
}

\newcommand{\phipropagator}[7]{
\draw [#6,ultra thick] (#1,#2) -- (#1/2+#3/2,#2/2+#4/2);
\draw [#7,ultra thick] ((#1/2+#3/2,#2/2+#4/2) -- (#3,#4);	
\node at (#1/2+#3/2+.5,#2/2+#4/2+.5) {$#5$};	
} 

\newcommand{\phipropagatorr}[5]{
\phipropagator{#1}{#2}{#3}{#4}{#5}{phir}{phir}
}

\newcommand{\phipropagatorp}[5]{
\phipropagator{#1}{#2}{#3}{#4}{#5}{phir}{phil}
} 

\newcommand{\phipropagatorm}[5]{
\phipropagator{#1}{#2}{#3}{#4}{#5}{phil}{phir}
}

\newcommand{\phipropagatorl }[5]{
\phipropagator{#1}{#2}{#3}{#4}{#5}{phil}{phil}
}

\newcommand{\drawphichidiagb}[7]{
\begin{scope}[shift={(#1,#2)}]
\begin{scope}[rotate=#3+180]
\draw [#4,ultra thick, domain=270:360] plot ({1*cos(\x)}, {1*sin(\x)});
\draw [#5, ultra thick, domain=180:270] plot ({1*cos(\x)}, {1*sin(\x)});
\draw [#6, ultra thick, domain=80:180] plot ({1*cos(\x)}, {1*sin(\x)});
\draw [#7, ultra thick, domain=0:90] plot ({1*cos(\x)}, {1*sin(\x)});
\node at (1,0) {$\times $};	
\node at (-1,0) {$\times $};		  
\end{scope}    
\end{scope}     
} 

\newcommand{\drawphichidiagbrr}[3]{
\drawphichidiagb{#1}{#2}{#3}{phir}{phir}{chir}{chir}
} 

\newcommand{\drawphichidiagbrp}[3]{
\drawphichidiagb{#1}{#2}{#3}{phir}{phir}{chir}{chil}
} 

\newcommand{\drawphichidiagbrm}[3]{
\drawphichidiagb{#1}{#2}{#3}{phir}{phir}{chil}{chir}
} 

\newcommand{\drawphichidiagbrl}[3]{
\drawphichidiagb{#1}{#2}{#3}{phir}{phir}{chil}{chil}
} 

\newcommand{\drawphichidiagbpr}[3]{
\drawphichidiagb{#1}{#2}{#3}{phir}{phil}{chir}{chir}
} 

\newcommand{\drawphichidiagbpp}[3]{
\drawphichidiagb{#1}{#2}{#3}{phir}{phil}{chir}{chil}
}

\newcommand{\drawphichidiagbpm}[3]{
\drawphichidiagb{#1}{#2}{#3}{phir}{phil}{chil}{chir}
}

\newcommand{\drawphichidiagbpl}[3]{
\drawphichidiagb{#1}{#2}{#3}{phir}{phil}{chil}{chil}
}

\newcommand{\drawphichidiagbmr}[3]{
\drawphichidiagb{#1}{#2}{#3}{phil}{phir}{chir}{chir}
}

\newcommand{\drawphichidiagbmp}[3]{
\drawphichidiagb{#1}{#2}{#3}{phil}{phir}{chir}{chil}
}

\newcommand{\drawphichidiagbmm}[3]{
\drawphichidiagb{#1}{#2}{#3}{phil}{phir}{chil}{chir}
}

\newcommand{\drawphichidiagbml}[3]{
\drawphichidiagb{#1}{#2}{#3}{phil}{phir}{chil}{chil}
}
 
\newcommand{\drawphichidiagblr}[3]{
\drawphichidiagb{#1}{#2}{#3}{phil}{phil}{chir}{chir}
}

\newcommand{\drawphichidiagblp}[3]{
\drawphichidiagb{#1}{#2}{#3}{phil}{phil}{chir}{chil}
}

\newcommand{\drawphichidiagblm}[3]{
\drawphichidiagb{#1}{#2}{#3}{phil}{phil}{chil}{chir}
}

\newcommand{\drawphichidiagbll}[3]{
\drawphichidiagb{#1}{#2}{#3}{phil}{phil}{chil}{chil}
}

\newcommand{\psipropagator}[7]{
\draw [#6, ultra thick][->] (#1,#2) -- (#1/2+#3/2,#2/2+#4/2);
\draw [#7,ultra thick] ((#1/2+#3/2,#2/2+#4/2) -- (#3,#4);	
\node at (#1/2+#3/2+.5,#2/2+#4/2+.5) {#5};	
} 

\newcommand{\psipropagatorr}[5]{
\psipropagator{#1}{#2}{#3}{#4}{#5}{psir}{psir}
} 
\newcommand{\psipropagatorm}[5]{
\psipropagator{#1}{#2}{#3}{#4}{#5}{psil}{psir}
}
\newcommand{\psipropagatorp}[5]{
\psipropagator{#1}{#2}{#3}{#4}{#5}{psir}{psil}
} 
\newcommand{\psipropagatorl}[5]{
\psipropagator{#1}{#2}{#3}{#4}{#5}{psil}{psil}
}

\newcommand{\psibarpropagator}[7]{
\draw [#6, ultra thick][->] (#3,#4) -- (#1/2+#3/2,#2/2+#4/2);
\draw [#7,ultra thick] ((#1,#2) -- (#1/2+#3/2,#2/2+#4/2);	
\node at (#1/2+#3/2+.5,#2/2+#4/2+.5) {#5};	
} 

\newcommand{\psibarpropagatorr}[5]{
\psibarpropagator{#1}{#2}{#3}{#4}{#5}{psir}{psir}
}

\newcommand{\psibarpropagatorl}[5]{
\psibarpropagator{#1}{#2}{#3}{#4}{#5}{psil}{psil}
}

\newcommand{\drawpsidiaga}[5]{
\begin{scope}[shift={(#1,#2)},rotate=#3]
\draw [#4, ultra thick, domain=180:360] plot ({1*cos(\x)}, {1*sin(\x)});
\draw [#5, ultra thick, domain=0:180][->] plot ({1*cos(\x)}, {1*sin(\x)});
\node at (1,0) {$\times $};	
\end{scope}    
}

\newcommand{\drawpsidiagar}[3]{
\drawpsidiaga{#1}{#2}{#3}{psir}{psir}
}

\newcommand{\drawpsidiagap}[3]{
\drawpsidiaga{#1}{#2}{#3}{psir}{psil}
}

\newcommand{\drawpsidiagam}[3]{
\drawpsidiaga{#1}{#2}{#3}{psil}{psir}
}

\newcommand{\drawpsidiagal}[3]{
\drawpsidiaga{#1}{#2}{#3}{psil}{psil}
}

\newcommand{\drawpsidiagb}[7]{
\begin{scope}[shift={(#1,#2)}]
\begin{scope}[rotate=#3]
\draw [#4,ultra thick, domain=270:360] plot ({1*cos(\x)}, {1*sin(\x)});
\draw [#5, ultra thick, domain=180:270][->] plot ({1*cos(\x)}, {1*sin(\x)});
\draw [#6, ultra thick, domain=80:180] plot ({1*cos(\x)}, {1*sin(\x)});
\draw [#7, ultra thick, domain=0:90][->] plot ({1*cos(\x)}, {1*sin(\x)});
\node at (1,0) {$\times $};	
\node at (-1,0) {$\times $};		 
\end{scope}    
\end{scope}    
} 

\newcommand{\drawpsidiagbrr}[3]{
\drawpsidiagb{#1}{#2}{#3}{psir}{psir}{psir}{psir}
}

\newcommand{\drawpsidiagbmr}[3]{
\drawpsidiagb{#1}{#2}{#3}{psir}{psil}{psir}{psir}
}  

\newcommand{\drawpsidiagbmm}[3]{
\drawpsidiagb{#1}{#2}{#3}{psir}{psil}{psir}{psil}
}

\newcommand{\drawpsidiagbmp}[3]{
\drawpsidiagb{#1}{#2}{#3}{psir}{psil}{psil}{psir}
}

\newcommand{\drawpsidiagbml}[3]{
\drawpsidiagb{#1}{#2}{#3}{psir}{psil}{psil}{psil}
}

\newcommand{\drawpsidiagbpr}[3]{
\drawpsidiagb{#1}{#2}{#3}{psil}{psir}{psir}{psir}
}

\newcommand{\drawpsidiagbpm}[3]{
\drawpsidiagb{#1}{#2}{#3}{psil}{psir}{psir}{psil}
}

\newcommand{\drawpsidiagbpp}[3]{
\drawpsidiagb{#1}{#2}{#3}{psil}{psir}{psil}{psir}
}

\newcommand{\drawpsidiagbpl}[3]{
\drawpsidiagb{#1}{#2}{#3}{psil}{psir}{psil}{psil}
}
 
\newcommand{\drawpsidiagblr}[3]{
\drawpsidiagb{#1}{#2}{#3}{psil}{psil}{psir}{psir}
}

\newcommand{\drawpsidiagbll}[3]{
\drawpsidiagb{#1}{#2}{#3}{psil}{psil}{psil}{psil}
}

\def\psibr{\bar{\psi}_{\textrm{R}}} 
\def\psibl{\bar{\psi}_{\textrm{L}}}

\def\psir{\psi_{\textrm{R}}} 
\def\psil{\psi_{\textrm{L}}}

\def\pgator{ \mathcal{P} }   
\def\mommea{ \int [\mathcal{D}p]  }  
\def\mommeasure{  [\mathcal{D}p]  }

\hbadness=\maxdimen
\vbadness=\maxdimen

\allowdisplaybreaks[1]

\title{\boldmath Renormalisation in Open Quantum Field theory II: Yukawa theory and PV reduction}

\author[\symbishop ,\symknight]{Avinash }
\author[\symking]{, Chandan Jana }
\author[\sympawn]{, Arnab Rudra }

\affiliation[\symbishop]{Indian Institute of Science,\\
C.V. Raman Avenue, Bangalore 560012, India.\\}
\affiliation[\symknight]{Department of Physics,\\
University of California, Davis, CA 95616 USA.\\}
\affiliation[\symking]{International Centre for Theoretical Sciences (ICTS-TIFR)\\ 
Shivakote, Hesaraghatta Hobli, Bengaluru 560089, India.\\}
\affiliation[\sympawn]{International Centre for Theoretical Physics\\ 
Strada Costiera 11, Trieste 34151 Italy.\\}
\emailAdd{baidyaavinash@gmail.com} 
\emailAdd{chandan.jana@icts.res.in} 
\emailAdd{rudra.arnab@gmail.com}  

\abstract{We compute Passarino-Veltman (PV) reduction for tensor loop integrals, that appear in open field theories. We apply these results to open-Yukawa theory and compute the self-energy correction of the fields. We found that non-local divergences show up in the one loop correction to the fermionic self-energy. These non-local divergences do not disappear even if the tree level theory is chosen to satisfy the trace preserving condition of the density matrix. }

\begin{document}
\maketitle
\raggedbottom

\newpage

\section{Introduction}
\label{sec:bjrintro}

Field theory is a framework to study systems where the particle number is not conserved and Quantum field theory provides a formalism to study quantum dynamics of such systems. For any system, quantum corrections often alter the nature of classical dynamics. For example, quantum corrections can make a classical vacuum unstable \cite{Coleman:1973jx}. Moreover, studies of anomalies and renormalization group provide us guiding principle to identify the space of Quantum field theories that can describe a physical system. That is why it is very important to study the quantum corrections of any field theory. In Quantum field theory, Feynman diagrams provide a diagrammatic way to organize the perturbative computations. In the language of Feynman diagrams, tree level diagrams capture the classical dynamics and the quantum effects are encoded in the loop amplitudes. The leading quantum correction is encoded in one loop amplitudes and often the study of one loop amplitudes is enough to understand the nature of quantum correction(s). For example, anomalies are captured entirely by one loop amplitudes.  The nature of renormalization group flow \footnote{Sign of the beta function.} near the fixed points is almost determined by one loop corrections. So clearly one loop corrections are extremely important in any Quantum field theory. 

The loop amplitudes are constructed out of propagators (of various fields) and the various interaction terms in the Lagrangian. However, the field content and the interactions differ in various quantum field theories. A priori, it seems that the study of loop amplitudes is a task that has to be executed separately in a case by case fashion for various theories. In 1979, Passarino and Veltman \cite{Passarino:1978jh}  came up with a framework to study one loop amplitudes up to four internal legs for any quantum field theory \footnote{They were primarily interested in four-dimensional quantum field theory. Their work also holds for higher dimension. }. They showed that any loop amplitude  (with up to four internal legs) can be written in term of four basic loop integrals constructed out of one, two, three and four scalar propagators \footnote{t' Hooft and Veltman evaluated these scalar integrals explicitly \cite{tHooft:1978jhc}. }. They provided an algorithm (which is referred to as Passarino-Veltman tensor reduction) to reduce any loop integrals in terms of the scalar integrals. Later this approach was extended to one loop amplitudes with more than four integral legs \cite{Denner:2002ii, Denner:2005nn}.  This method is extremely useful to study one-loop corrections in four-dimensional quantum field theories. We still do not know the extension of this method of tensor reduction for two and higher loop diagrams.  In this paper, we explore the extension of this method in a different direction. We apply the method of tensor reduction to Schwinger-Keldysh (henceforth SK) field theories and then we apply it to Yukawa theory on SK contour. SK theory is a framework to compute correlation functiozn for a system with no-prior information of its final state \cite{Keldysh:1964ud, Schwinger:1960qe} (See \cite{Haehl:2016pec, Haehl:2016uah} for a recent review of SK formalism). 

In high-energy physics, we mostly study ``in-out'' correlators. These ``in-out" correlators are related to the $S$ matrices which are the primary observables in a high energy experiment.   We assume that the initial and the final vacuum state of the interacting theory is same up to a phase and we compute the correlator between in vacuum and out vacuum. However, in many cases (for example, systems not in equilibrium, ...) the initial state and the final state are not the same up to a phase. In fact, sometimes we do not have any prior knowledge of the final state. The SK formalism provides a framework to study quantum dynamics without assuming anything about the final state. This formalism is based on a closed time contour and it is extremely useful to study the time evolution of mixed states, which arise in open quantum systems, out-of-time-ordered correlators, non-equilibrium dynamics etc. \cite{Kamenev, Chou:1984es}.

Though the SK formalism is old, not much is known about the loop corrections in this formalism. In particular, the most general (local) field theory that one can write down in SK contour is not unitary. They describe open quantum systems. Amongst these non-unitary theories, there is a sub-class of theories in which the time evolution preserves the trace of the density matrix. We call such theories as {\it Lindblad  theories} \cite{Gorini:1975nb, Lindblad:1975ef}.   In unitary theory, the trace of any moment of the density matrix (i.e. $\rho^n $ for $n\geq 1$) is preserved under time evolution. This is not necessarily true for Lindblad theories. 
  
Lindblad theories were originally developed to describe non-relativistic Markovian open quantum mechanical systems. They have been useful to describe stochastic, non-equilibrium quantum systems. One can easily extend these quantum mechanical theories to field theories by demanding the theory to be local in both space and time \cite{Avinash:2017asn}. The connection of these theories with an underlying unitary theory is not yet clear. One expects that the theory obtained by tracing out some light degrees of freedom from a unitary QFT to be Lindblad theory and the unitarity of the microscopic theory implies the Lindblad condition for the open EFT.

The simplest of such case which consists of only a real self-interacting scalar in the SK contour was studied in \cite{Avinash:2017asn} where it was shown that the Lindblad conditions are preserved under renormalization group flow. The next obvious step is to study more general Lindblad theories with fermions, vector bosons etc. The results in this paper is a step towards that direction. In order to study one loop beta function in more general field theory, we extend PV tools to more general theories on the SK contour. Then we apply those results to compute mass renormalization in open Yukawa theory. 
 
 The organization of the paper is as follows. First, we discuss the program of renormalisation in open-field theories in section \S\ref{sec:bjrrenormsk}. In section \S\ref{sec:pvebasics}, we perform the PV reduction for open QFTs. We start by reviewing the basics of PV reduction in unitary theories. This also serves the purpose to introduce various (standard) notations. We, then, generalise the idea of PV reduction to SK theory and analyse various $A$-type and $B$-type integrals (explained later) in SK theory. In section \S\ref{sec:bjropenyukawa}, we implement the PV reductions in open-Yukawa theory and compute the self-energy correction to the fields. The appendices complement the computations in the main body. In appendix \ref{sec:feynmandiagram} and in appendix  \ref{sec:fermionicpv} we draw the diagrams used in the self-energy correction to the fields and write the PV reduction for diagrams used in fermionic self-energy correction respectively. In appendix \ref{app:twoscalar} we discuss non-local divergences that appear in one loop correction to the quartic couplings of two scalars.


\section{Renormalization of open quantum field theories} 
\label{sec:bjrrenormsk}

Unitary field theories describe closed systems. Renormalization in unitary theories provides us a guiding principle to write down models to describe various systems.  
As we have described in the introduction, the SK formalism can provide a description for the open systems. If one writes down the most general field theory on the SK contour then those theories are not unitary. Only a subclass of these non-unitary theories preserve the trace of the density matrix. 

The space of relativistic open quantum field theories is mostly unexplored. In \cite{Avinash:2017asn}, the authors started a program to understand the space of open relativistic EFTs.   We summarize the key points of that programme here.
\begin{enumerate} 

\item  One key assumption is that there exists open quantum systems that are described by {\it local} quantum field theories. The hope was to derive at least one such local theory from an underlying unitary field theory using the Feynman-Vernon method \cite{Feynman:1963fq}. But this idea has not materialized yet \cite{ Chatterjee:2019xxx}.

Among these local field theories only in the Lindblad theories the time-evolution preserves the trace of the density matrix.

    \item Even if the derivation of local open quantum field theories is still missing, the hope is to explore the space of such open quantum field theories in the Wilsonian approach. In particular, the goal is to explore whether 1) the criteria of the locality of the theory, 2) the criteria of renormalizability and 3) the Lindblad conditions \footnote{This is the closest analogue of unitarity in open QFT.} are mutually consistent with each other. And if they are consistent then what is the space of such theories? For example, One could ask the following questions.
\begin{itemize}
             
\item whether the Lindblad theories (and/or more general theories in the SK contour) are renormalizable.

\item what are the criteria for renormalizability ?
 
\item how are the beta functions affected by non-unitarity dynamics ?

\end{itemize}

\end{enumerate}
In the following subsection, we discuss the progresses that has been made to address these questions with the above-mentioned assumptions.

\subsection{Various examples and failure}
\label{subsec:bjrrenormskexample}

The first step in this direction was taken in \cite{Avinash:2017asn} where 
the authors considered open $\phi^3+\phi^4$ theory. At first, the most general theory of a single real scalar field, in the SK contour, was written down. It was found that the Lindblad conditions are protected at one loop. Moreover, it was found that these subclass of theories are renormalizable and the trace-preserving conditions \&  the renormalizability condition are mutually consistent. 

The next simplest example is an open field theory of two scalars ($\phi,\,\chi$) with a $\phi^2\, \chi^2$ interaction term. The details of this model can be found in appendix \ref{app:twoscalar}. The masses of the scalars are chosen to be different ($m_1,\,m_2$). It has been shown in \cite{Avinash:2017asn} that few bubble loop diagrams with two internal propagators of different fields have non-local divergence structure\footnote{The authors of \cite{Burgess:2018sou} considered QFT for Rindler observers and they found UV-IR mixing in loop corrections to the correlators \& the breakdown of perturbative expansion.}. One such integral \footnote{This divergence come from an integral, named $B_{RP}(m_1,m_2;k)$. The rule behind this naming is explained in section \S\ref{sec:pvebasics}.} has the following divergence structure. 
\begin{equation}
\begin{split}
\label{divstructure}
\frac{i}{(4\pi)^2} \frac{k^2-m_1^2+m_2^2}{2k^2}\, \frac{2}{d-4}\,.
\end{split}
\end{equation}
This integral appear, for example, in one loop renormalization to $\phi_R^2\, \chi_R^2$ vertex (see section \ref{subsec:phichivertexcorrection}) and the non-local divergences don't cancel even if the tree level theory satisfies the trace-preserving/Lindblad conditions. The non-local divergence disappears in the equal mass limit.  The existence of the non-local divergences was a hint that these divergences would show up in more general field theories. However, there was a hope that in open theories, these divergences may be proportional to the Lindblad violating coupling. So, in Lindblad theories, they may vanish. Any local Lindblad theory would be renormalizable; these non-local divergences will not pose any threat to the Lindblad theories. In this particular case, we perform the explicit computations and we found that these divergences survive even in Lindblad EFT.   We do not present the computations of the two scalar theory in details here. Rather we present open Yukawa theory in length. The conclusion remains the same. Open Yukawa theory is also plagued by non-local divergences (even if the tree level dynamics preserves the trace of the density matrix). 

The structure of the non-local divergence in \eqref{divstructure} naively indicates that if all the particles have same mass then these divergences vanish. This expectation is only true in theories with scalars only (and even in that case, only upto diagrams with one and two internal propagators; For three or more internal propagators, there are non-local divergences \cite{ Chatterjee:2019xxx}). For theories with fermions and/or with vector bosons, the non-local divergences persist even in the equal mass limit. We illustrate this point explicitly by performing the PV reduction (see eqn \eqref{zero_gamma5_FS no cut equal mass} and eqn  \eqref{zero_gamma5_FS one cut equal mass}). 

We also have considered supersymmetric open Wess-Zumino theory and found that the non-local divergences remain \cite{Baidya:2019ab} .


\section{PV reduction in open QFTs }
\label{sec:pvebasics}

In this section, first we briefly review the method of PV tensor reduction. Simultaneously, we also introduce various notations which will be useful to extend the method of tensor reduction to SK theory.  In the standard literature, one loop amplitude with one, two, three and four internal propagators are referred as PV $A$-type, $B$-type, $C$-type and $D$-type integrals. We use the same terminology extensively. Passarino and Veltman considered integrals up to four internal propagators \cite{Passarino:1978jh, tHooft:1978jhc}. Later the case of one loop integral with more internal propagator was considered in \cite{Denner:2002ii, Denner:2005nn}. In our work, we considered integrals with less than four internal propagators and as a consequence, we also limit our discussion of the original work up to two propagators.  Consider the expression for tadpole, bubble and  triangle  diagram in $\phi^3$ theory \cite{tHooft:1978jhc}

\begin{eqnarray}
    A&=&
     \mommea
     \Bigg[
\frac{-i }{[p^2+m_0^2-i\varepsilon]}
\Bigg]
\quad, 
\\
    B&=&
     \mommea
     \Bigg[
\frac{(-i)^2 }{[p^2+m_0^2-i\varepsilon][(k_1+p)^2+m_1^2-i\varepsilon]}
\Bigg]
\quad, 
\\
    C&=&
     \mommea
     \Bigg[
\frac{(-i)^3 }{[p^2+m_0^2-i\varepsilon][(k_1+p)^2+m_1^2-i\varepsilon][(k_1+k_2+p)^2+m_2^2-i\varepsilon]}
\Bigg]
\quad. 
\end{eqnarray}
Here $\mommeasure$ is the measure for the loop integral
\begin{eqnarray}
\mommea =  \mu^{4-d} 
\int \frac{d^dp}{(2\pi)^d}\,
\quad.  
\end{eqnarray}
Passarino and Veltman showed that these integrals serve as the basis for any one loop integral with less than three internal legs. In $\phi^3$ scalar field theory, the numerator is very simple. But in a generic theory the numerator is a polynomial of internal and external momenta. Any polynomial which is a function of only the external momenta do not participate in the loop integration and hence they can taken out of the integral. Keeping this in mind, consider the following integral 
\begin{eqnarray}
&&
T^{\mu_1\dots\mu_q}_{r}(m_1,\dots ,m_{r}|k_1,\dots ,k_{r-1})
\label{pbvebasic1}
\\
&=& \mommea
\Bigg[
\frac{(-i)^r (p^{\mu_1}\dots p^{\mu_q})}{[p^2+m_1^2-i\varepsilon]
\dots 
[(k_1+\dots +k_{r-1}+p)^2+m_{r}^2-i\varepsilon]}
\Bigg]
\quad. 
\nonumber
\end{eqnarray} 
Here we explain various notations that are used in the above expression. 
\begin{itemize}
    \item $\mu_i$ are Lorentz indices - so they can take values from $0$ to $3$,

    \item $k_i^\mu $ s are external momenta,
    \item $p^\mu $ is the loop momentum ,
    \item $m_i$ is the mass of the $i$-th (internal-)propagator in the loop, 
    \item $r$ is the number of the external legs,
    \item $q$ is the number of Lorentz indices of the integral. It is not necessarily related to the number of the external legs. However, for a renormalizable unitary quantum field theory in $D$ dimensions $q\leq 2r- D$. 
\end{itemize}

Consider the collection of such integrals 
\begin{eqnarray}
    \{T^{\mu_1\dots \mu_q}_{r}\}\qquad,\qquad 1 \leq r \leq 4\qquad,\qquad 1 \leq q \leq r
\quad. 
\label{pbvebasic2}
\end{eqnarray}
Any one loop amplitude with at most 4 internal legs \footnote{In the present work, we restrict our attention to loop integrals with upto 3 internal legs.} in a generic quantum field theory can be written in terms of a linear combination of  $\{T^{\mu_1\cdots \mu_q}_{r}\}$s.  We elaborate this statement with one example. It is straightforward to show that  \citep{Bardin:1999ak}
\begin{eqnarray}
B^{\mu}(m_1,m_2,k) = \frac{k^\mu}{2k^2}\Big[- i A(m_2)+i A(m_1)+(k^2-m_1^2+m_2^2)\ B(m_1,m_2,k)\Big]\,\,.
\end{eqnarray}
For future convenience, we introduce a few more notations here. We introduce $\pgator (p,m)$ which just denotes the propagator for a scalar field with mass $m$ and momentum $p$ along with the Feynman $i \varepsilon$ prescription.  
\begin{eqnarray}
    \pgator (p,m)  &=&     \frac{-i}{p^2+m^2-i\varepsilon}
\quad. 
\label{pbvebasic3}
\end{eqnarray}  
Using this notation, we rewrite \eqref{pbvebasic1} 
\begin{eqnarray}
\mommea  
\,
(p^{\mu_1}\dots p^{\mu_q})
\pgator_{a_1} (p,m_1)
\dots 
\pgator_{a_{r}} ((k_1+\dots +k_{r-1}+p),m_{r})
\quad. 
\label{pbvebasic4}
\end{eqnarray}
The integral has $r$ Lorentz indices. The only available Lorentz tensors are $\eta^{\mu \nu}$ and $\{k_i^\mu\}$. So the integral can only be degree $r$ polynomial of these quantities times some Lorentz scalar.  For example, consider the integral $B^{\mu \nu}$. From Lorentz invariance it follows that it must be of the following form
\begin{eqnarray}
B^{\mu \nu }(k| m_1,m_2)&=&k^\mu k^\nu  B_{11}    (k| m_1,m_2)+\eta^{\mu \nu}  B_{00}    (k| m_1,m_2)
\quad. 
\label{pbvebasic5}
\end{eqnarray}
$B_{11}    (k| m_0,m_1)$ and $B_{00}    (k| m_0,m_1)$ are Lorentz scalars and they can be written in terms of PV $A$ and $B$ type integrals. We can write any tensor integral \footnote{In $D=4$, at most four vectors can be linearly independent of each other \cite{Denner:2002ii, Denner:2005nn}. }  in a similar way 
\begin{eqnarray}
T^{\mu_1\dots \mu_q}_{r} = k_{i_1}^{\mu_1} \dots k_{i_q}^{\mu_q}\,     T_{1}+ \dots 
\qquad.
\end{eqnarray}
As long as $k_i^{\mu}$s are linearly independent, this relation is invertible. The number of independent external  momenta is $q-1$ due to momentum conservation. 

The method to find explicit expressions for the scalar coefficient of these tensor structures is known as the method of tensor reduction. This method simplifies the numerator to get back the integrals of scalar $\phi^3$ theory. For example, multiplying $T^{\mu_1\cdots  \mu_q}_{r}$ by $k_1^\mu $ and using simple algebra, we get 
\begin{eqnarray}
2\,  k_1\cdot p= (k_1+p)^2-p^2= [(k_1+p)^2-m_1^2]-[p^2-m_0^2]+[m_1^2-m_0^2]
\quad.
\label{pbvebasic6}
\end{eqnarray}
Both the first and the second factor cancel one of the propagators. 

Using this we notice that the polynomial in the numerator has one less factor of $p^\mu$ 
\begin{eqnarray}
&& \mommea  
\,
(p^{\mu_2}\dots\, p^{\mu_q})
\pgator_{a_1} (p,m_1)
\dots 
\pgator_{a_{r}} ((k_1+\dots +k_{r-1}+p),m_{r})
\times 
\Big[(k_1+p)^2+m^2
\Big]
\quad.\qquad 
\label{pbvebasic7}
\end{eqnarray}
This concludes our brief review of PV tensor reduction for unitary theory. Now we move onto SK theory.

\subsubsection{$SK$ theory - Notation and convention}
\label{sec:pveskbasics}
\begin{figure}
\begin{center}
\begin{tikzpicture}

\draw[blue,ultra thick] (5,0) -- (9,0);
\node at (7,0.3) {$p\, \rightarrow$};
\node at (0,0) {$\pgator_R (p,m)$};
\node at (2,0) {$\textbf{:}$};

\draw[blue,ultra thick,dashed] (5,-1) -- (9,-1);
\node at (7,-0.7) {$p\, \rightarrow$};
\node at (0,-1) {$\pgator_L (p,m)$};
\node at (2,-1) {$\textbf{:}$};

\draw[blue,ultra thick] (5,-2) -- (7,-2);
\draw[blue,ultra thick,dashed] (7,-2) -- (9,-2);
\node at (7,-1.7) {$p\, \rightarrow$};
\node at (0,-2) {$\pgator_M (p,m)$};
\node at (2,-2) {$\textbf{:}$};

\draw[blue,ultra thick,dashed] (5,-3) -- (7,-3);
\draw[blue,ultra thick] (7,-3) -- (9,-3);
\node at (7,-2.7) {$p\, \rightarrow$};
\node at (0,-3) {$\pgator_L (p,m)$};
\node at (2,-3) {$\textbf{:}$};

\end{tikzpicture}
\end{center}
\caption{SK propagators}
\label{dia:propagators}
\end{figure}
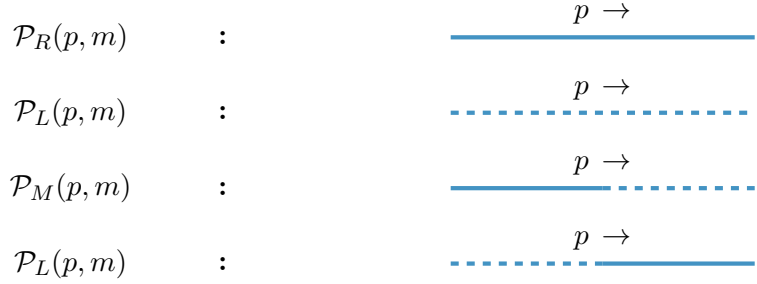

In $SK$ theory there are four type of propagators 
\begin{subequations}
\begin{eqnarray}
\pgator_R (p,m)  &=&     \frac{-i}{p^2+m^2-i\varepsilon}
\quad,
\label{pveskbasic1a}
\\
\pgator_P (p,m)  &=&     2\pi \Theta(p_0)\, \delta(p^2+m^2)\equiv 2\pi \delta_{+}(p^2+m^2)
\quad,
\label{pveskbasic1b}
\\
\pgator_M (p,m)  &=&     2\pi \Theta(-p_0)\, \delta(p^2+m^2)\equiv 2\pi\delta_{-}(p^2+m^2) 
\quad,
\label{pveskbasic1c}
\\
\pgator_L (p,m)  &=&     \frac{i}{p^2+m^2+i\varepsilon}
\quad. 
\label{pveskbasic1d}
\end{eqnarray} 
\end{subequations}
The $R$ and $L$ propagators are the time ordered and the anti-time ordered propagators respectively. The other two propagators are essentially on-shell propagators. These two propagators also constrain the flow of energy; $P$ propagator allows the positive frequencies to flow, whereas $M$ propagator allows negative frequencies. In fig. \ref{dia:propagators} we introduce the diagrammatic representation of the above propagators which we use later for $SK$ Feynman diagrams.

The propagators are not all independent, but related by the \emph{largest time equation} \cite{Cutkosky:1960sp, Veltman:1963th} which is also known to as \emph{cutting equation};
\begin{eqnarray}
    \textrm{Cutting equation}: \qquad\qquad \pgator_R (p,m) +\pgator_L (p,m) = \pgator_P (p,m)+\pgator_M (p,m) 
\quad.
\label{pveskbasic3}
\end{eqnarray}
Various SK propagators are also related by CPT.
\begin{eqnarray}
    \textrm{CPT}: \qquad\qquad \begin{matrix} 
  \pgator_R (p,m) \longleftrightarrow \pgator_L (p,m)\quad,  \\
\label{pveskbasic2}
  \\
  \pgator_M (p,m) \longleftrightarrow \pgator_P (p,m) \quad. 
\end{matrix}
\end{eqnarray}

In any SK field theory there are four propagators. It is useful to generalise the form of the loop integrals given in \eqref{pbvebasic1}, to SK theory in the following way 
\begin{eqnarray}
&&
T^{\mu_1 \dots \mu_q}_{a_1 \dots  a_r}(m_1,\dots ,m_{r}|k_1,\dots , k_{r-1})
\label{general_SK_diagram}
\\
&=&\mommea \,
(p^{\mu_1}\dots p^{\mu_q})
\pgator_{a_1} (p,m_1)\, 
\pgator_{a_2} (k_1+p,m_2)
\dots 
\pgator_{a_{r}} ((k_1+\dots+k_{r-1}+p),m_{r})
\quad. 
\nonumber
\end{eqnarray}  
Compared to equation \eqref{pbvebasic4} here the only new additions are the subscripts $a_i$s - they denote the type of propagators. They take values among - $R,P,M,L$. The naive counting implies that, at any order, the number of integrals in SK theory is $4^r$ times the number of integrals in unitary theory (Here $r$ is the number of internal legs).  However, just by using CPT (given in equation \eqref{pveskbasic2}) and cutting identity(given in equation \eqref{pveskbasic3}) we can reduce the number of independent integrals.  However, it's worth mentioning that  by using \eqref{pveskbasic2}  and \eqref{pveskbasic3}, all SK loop integrals cannot be written entirely in terms of the loop integrals of unitary theory; there are genuine SK integrals which are not present in an unitary theory. 


\paragraph{Convention for the direction of loop momenta:}

The propagators of a unitary theory depend only on the magnitude of the momentum, not on their direction.  But, the $P$ and $M$ propagators in fig \ref{dia:propagators} depend on the direction of the momenta. If the direction of momentum is flipped then the $P$ and $M$ propagator interchange. The subscript of an SK integral (as denoted in equation \eqref{general_SK_diagram}) can change depending on the flow of momenta. We adopt a convention such that there is a one-one correspondence between a subscript and a one loop SK diagrams. The convention is the following. In the definition of most general SK integral in \eqref{general_SK_diagram} we assumed that the leftmost subscript ($a_1$) in $T^{\mu_1 \dots \mu_q}_{a_1 \dots  a_r}$ corresponds to the propagator which has no dependence on external momenta. Staring from this propagator the rest of the propagators are drawn in a counter-clockwise (C.C.W.) direction which is in one to one correspondence with the subscript labels in $T^{\mu_1 \dots \mu_q}_{a_1 \dots  a_r}$. To explain this, let us consider the following examples.

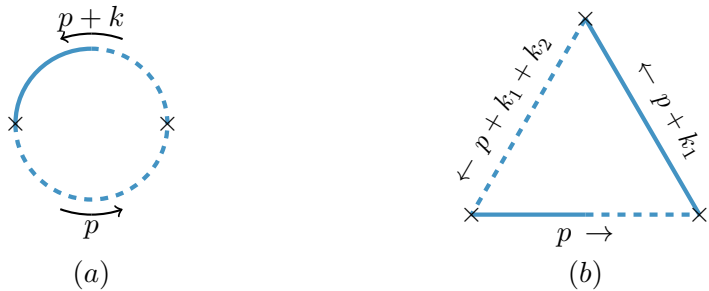
\begin{figure} 
\begin{center}
\begin{tikzpicture}

\begin{scope}[shift={(0,0)}]
\draw[blue,ultra thick,dashed] (-1,0) arc (180:360:1);
\draw[blue,ultra thick,dashed] (1,0) arc (0:90:1);
\draw[blue,ultra thick] (0,1) arc (90:180:1);

\draw[thick,->] (0,-1.2) arc (270:290:1.2);
\draw[thick] (0,-1.2) arc (270:250:1.2);
\node at (0,-1.4) {$p$};

\draw[thick,->] (0,1.2) arc (90:110:1.2);
\draw[thick] (0,1.2) arc (90:70:1.2);
\node at (0,1.4) {$p+k$};

\node at (-1,0) {$\times$};
\node at (1,0) {$\times$};

\node at (0,-2) {$(a)$};
   
\end{scope} 

\begin{scope}[shift={(5,-1.2)}]

\coordinate(A) at (0,0);
\coordinate(D) at (1.5,0);
\coordinate(B) at (3,0);
\coordinate(C) at ($(A)!1!60:(B)$);

\draw[blue,ultra thick] (A) -- (D);
\draw[blue,ultra thick,dashed] (D) -- (B);
\draw[blue,ultra thick] (B) -- (C);
\draw[blue,ultra thick,dashed] (C) -- (A);

\node at (A) {$\times$};
\node at (B) {$\times$};
\node at (C) {$\times$};

\node at (1.5,-.3) {$p\, \rightarrow$};
\node[rotate=-60] at (2.6,1.4) {\small{$\leftarrow\, p+k_1$}};
\node[rotate=60] at (0.4,1.4) {\small{$\leftarrow\, p+k_1+k_2$}};
\node at (1.5,-.8) {$(b)$};
\end{scope}

\end{tikzpicture}
\end{center}
\caption{These diagrams are schematic representations of the corresponding SK integrals. $(a)$ correspond to $B^\mu_{LM}$ and $(b)$ correspond to $C^{\mu\nu}_{PRL}$}.
\label{dia:BMLandCPRL}
\end{figure}

Let us choose $B^\mu_{LM}(m_1,m_2;k)$ from $B$ type integrals and $C^{\mu\nu}_{PRL}(m_1,m_2,m_3;k_1,k_2)$ from $C$ type integrals. The diagrammatic expressions are shown in fig. \ref{dia:BMLandCPRL}. 
For $B^\mu_{LM}$, following our convention,  we start from the $L$ propagator which has no external momentum dependence. The next propagator can either be $P$ or $M$ depending on the direction of momentum. But our rule for the direction of momenta is to follow C. C. W., which eliminates the ambiguity. The same logic also applies to $C^{\mu\nu}_{PRL}$. It is straightforward to check that the diagram $(b)$ in fig. \ref{dia:BMLandCPRL} for $C^{\mu\nu}_{PRL}$ is consistent with our convention.


\subsection{$A$ type integrals}
\label{sec:pveskatype}
In this section, we consider the $A$ type integrals in the SK theories. In a renormalizable theory, these integrals can come in one loop correction to the one point function in any theory with three-point contact interaction(s) (for example $\phi^3$ theory, Yukawa theory) and/or to the one loop correction of the self-energy in any theory with four-point contact interaction(s) (for example $\phi^4$ theory). The $A$ type integrals in open $\phi^3+\phi^4$ theory was computed in \cite{Avinash:2017asn}. There are four scalar $A$ type SK integrals. Those are given by,
\begin{eqnarray}
A_R\quad,\quad A_L \quad,\quad A_P\quad,\quad A_M
\quad.  
\label{scalaratype1}
\end{eqnarray}
In this case, the integrals are simple and one can explicitly evaluate all of them.
\begin{eqnarray}
    A_R&=&\mu^{4-d}\int \frac{d^dp}{(2\pi)^d} \frac{-i}{p^2+m^2-i\varepsilon}
\quad, 
\label{scalaratype2a}
\\
    A_L&=&\mu^{4-d}\int \frac{d^dp}{(2\pi)^d} \frac{i}{p^2+m^2+i\varepsilon}
\quad, 
\label{scalaratype2b}
\\
    A_P&=&\mu^{4-d}\int \frac{d^dp}{(2\pi)^d}\,  2\pi\, \delta_+\left((p-k)^2+m_2^2\right)
\quad, 
\label{scalaratype2c}
\\
    A_M&=&\mu^{4-d}\int \frac{d^dp}{(2\pi)^d}\,  2\pi\, \delta_-\left((p-k)^2+m_2^2\right)
\quad. 
\label{scalaratype2d}
\end{eqnarray}
But we have mentioned that one can use eqn  \eqref{pveskbasic2} and eqn \eqref{pveskbasic3} to reduce the number of independent integrals. We  demonstrate it explicitly in this simple case of $A$ type integrals \cite{Avinash:2017asn}.   The action of CPT implies that 
\begin{eqnarray}
     A_R=A_L \qquad, \qquad A_P=A_M     
\quad.  
\label{scalaratype3}
\end{eqnarray}
In this case, we do not get any new relation using the cutting identity (given in equation \eqref{pveskbasic3}). So naively the number of independent integral is $2$. We can explicitly evaluate them and we find that all of the integrals are same.
\begin{eqnarray}
A_R=\frac{m^2}{(4\pi)^2}\left[\frac{2}{d-4}+\ln \frac{m^2}{4\pi \mu^2 e^{-\gamma_E}}-1 \right]=A_P    
\quad.  
\label{scalaratype4} 
\end{eqnarray}

\subsubsection{Vector integrals $A^{\mu}_{a}(m)$}
The discussion of $A$ type vector integrals involve four integrals. First we discuss the following integral (which can appear in a unitary theory).  
\begin{equation}
\begin{split}
A^{\mu}_{R}(m) = \int \frac{d^dp}{(2\pi)^d} \frac{-i\,  p^\mu}{p^2+m^2-i\varepsilon}
\end{split}
\quad.  
\label{vectoratype1}
\end{equation}
This equation is odd in the integration variable and hence is zero. This holds true for $A^{\mu}_{L}(m)$ but does not hold for $A^{\mu}_{P}(m)$ and $A^{\mu}_{M}(m)$. This is because of the fact that both these integrals have a step function in the time-like component of the momentum.   But the sum of these two is given by 
\begin{equation}
\begin{split}
A^{\mu}_{P}(m)+A^{\mu}_M(m) =\int \frac{d^dp}{(2\pi)^d} \ p^\mu\ \delta(p^2+m^2) = 0 
\end{split}
\quad.  
\label{vectoratype2}
\end{equation} 
 The interesting fact is that CPT implies that these two integrals always appear simultaneously with the same coupling constant. So we can always add these two loop integrals and total contribution $0$.

\subsection{$B$ type integrals}
\label{sec:pveskbtype}

We consider $B$ type diagrams - integrals with two propagators. There are sixteen SK $B$-type integrals. In this section, we discuss vector and tensor $B$-type one loop integrals (up to 2-vector indices) that can appear in open QFTs. We start by reviewing the $B$ type scalar integrals (For elaborate discussion check appendix $B$ of \cite{Avinash:2017asn}).

\subsubsection{Scalar integrals $B_{ab}\, (m_1,m_2)$}
\label{subsec:pveskbtypescalar}

The  sixteen $B$-type scalar integrals are listed below 
\begin{equation}
\label{bscalar1}
\begin{split}
B_{RR}, \ \ \ B_{LL}, \ \ \ {B_{RL}}, \ \ \ B_{LR},\\
B_{PP}, \ \ \ B_{MM}, \ \ \ {B_{PM}}, \ \ \ B_{MP},\\
B_{RP}, \ \ \ B_{RM}, \ \ \ {B_{PR}}, \ \ \ B_{MR},\\
B_{LP}, \ \ \ B_{LM}, \ \ \ {B_{PL}}, \ \ \ B_{ML}.\\
\end{split}
\end{equation}
As already described, here $R,\,L$ correspond to propagators in the SK contour. One can write the integral expressions of these loops using the rules explained in section (\S\ref{sec:pveskbasics}).
These loop integrals were computed in \cite{Avinash:2017asn} and we only review the main results here.  
\begin{itemize}
    \item One can show that all these integrals are not independent using the CPT symmetry (eqn \eqref{pveskbasic2}) and the cutting relation (eqn  \eqref{pveskbasic3}). In fact all of these loop integrals can be expressed in terms of one master loop integral, $B_{RP}$. Let us write few relations, e.g.,
\begin{equation}
B_{RM}(k,m_1,m_2) = B_{RP}(-k,m_1,m_2) \qquad,\qquad  B_{LP}(k,m_1,m_2) = [B_{RP}(k,m_1,m_2)]^\ast \qquad .  
\label{bscalar2}
\end{equation}
The rest of the relations of $B$ type scalar loop integrals with $B_{RP}$ can be found in \cite{Avinash:2017asn}.
\item The loops in the second row of \eqref{bscalar1} are convergent when computed in dimensional regularisation. $B_{RL}$ and $B_{LR}$ are convergent if the internal propagators carry equal mass ($m_1=m_2$). The rest of the integrals are divergent in the equal mass limit\footnote{We studied only the case where $m_1= m_2$ in \citep{Avinash:2017asn}. The $m_1 \neq m_2$ case is not well understood as yet. But to keep our discussion on PV reduction general, we keep the masses to be unequal in all loop diagrams.}.
\item The divergence of diagrams in the third row of (\ref{bscalar1}) is half of the divergences of $B_{RR}$ and the divergences of the diagrams in the fourth row are half of the divergences of $B_{LL}$ in the equal mass limit.
\end{itemize} 
One can read off the divergence structure of all $B$ type scalar integrals from table (\ref{Tab:divergences}).
\begin{center}\label{Tab:divergences}
\begin{tabular}{|c|| c | c | c | c |}
\hline
 & R & L & P & M \\
\hline
R & $\Upsilon_{RR}$ & $\Upsilon_{RL}$ & $\frac{1}{2}(\Upsilon_{RR}+\Upsilon_{RL})$ & $\frac{1}{2}(\Upsilon_{RR}+\Upsilon_{RL})$\\
\hline
L & $-\Upsilon_{RL}$ & $-\Upsilon_{RR}$ & $-\frac{1}{2}(\Upsilon_{RR}+\Upsilon_{RL})$ & $-\frac{1}{2}(\Upsilon_{RR}+\Upsilon_{RL})$\\
\hline
P & $\frac{1}{2}(\Upsilon_{RR}-\Upsilon_{RL})$ & $-\frac{1}{2}(\Upsilon_{RR}-\Upsilon_{RL})$ & 0 & 0\\
\hline
M & $\frac{1}{2}(\Upsilon_{RR}-\Upsilon_{RL})$ & $-\frac{1}{2}(\Upsilon_{RR}-\Upsilon_{RL})$ & 0 & 0\\
\hline
\end{tabular}
\end{center}
where $\Upsilon_{RR}$ and $\Upsilon_{RL}$ in $\msbar$ are given by,
\begin{eqnarray}
\Upsilon_{RR} &=&\frac{2i}{(4\pi)^2}\left[\frac{1}{d-4} + \frac{1}{2}(\gamma_E-1-\ln\, 4\pi)\right]
\quad, 
\label{bscalar3a}
\\
\Upsilon_{RL} &=&\frac{2i}{(4\pi)^2}\frac{m_1^2-m_2^2}{k^2}\left[\frac{1}{d-4} + \frac{1}{2}(\gamma_E-1-\ln \, 4\pi)\right]
\quad.
\label{bscalar3b}
\end{eqnarray}
With this information of the scalar $B$ type loop integrals, we start discussing the PV reduction of the $B$ type tensor loop integrals.

\subsubsection{$B^{\mu}_{ab}(m_1,m_2)$ vector integrals}
\label{subsec:pveskbtypevector}

We show that $B^{\mu}_{ab}(m_1,m_2)$ can be expressed in terms of $A$ and $B$ type scalar loop integrals, as adversied in section \S\ref{sec:pvebasics}. These kind of loop integrals appear in the self energy correction to fermion. The vector integral is of the following form.
\begin{eqnarray}
B^{\mu}_{ab}(m_1,m_2;k) =
\mommea  
\,
(p^{\mu})
\pgator_{a} (p,m_1)
\pgator_{b} ((k+p),m_{2})
\label{vectorbtype0}
\end{eqnarray}
Again there are total sixteen $B^\mu$ integrals. The strategy that we follow in this section, is the following. Each $B^{\mu}$ type diagram is a Lorentz vector. So final answer must be a Lorentz vector. But the only vector present in $B^\mu$ type integrals is the external momentum $k^\mu$ \cite{Passarino:1978jh}. So we can write 
\begin{equation}
B^{\mu}_{ab}(m_1,m_2,k) \equiv k^\mu B^{(1)}_{ab} \qquad ,\qquad  \forall \quad a,b \in \{R,L,P,M\}
\label{vectorbtype1}
\end{equation} 
here $B^{(1)}_{ab}$ is a Lorentz scalar which is the proportionality constant. We determine it in terms of $B_{ab}$ and $A_{a}$. In order to do so, we  multiply both sides of \eqref{vectorbtype1} by $k_\mu$. This  give us the expression for $B^{(1)}_{ab}$
\begin{eqnarray}
k^2\, B^{(1)}_{ab}= k_\mu\,     B^{\mu}_{ab}(m_1,m_2,k) 
\quad. 
\label{vectorbtype2}
\end{eqnarray}
Now we implement this strategy for all the sixteen integrals. We organize our discussion based on the number of cut propagators. All sixteen integrals can be arranged into three different classes based on the number of cut propagators present in the integrals. First we discuss the integrals with no cut propagator.

\subsubsection*{Integrals with no cut propagators}
\label{subsubsec:pveskbtypenocutvector}

If we have no cut propagator (i.e. no $P$ and $M$ propagator) then the diagrams are made out of only $R$ and $L$ propagators. There are four such possibilities. They are given by
\begin{equation}
\label{vectorbtype3}
\begin{split}
&B^{\mu}_{RR}(m_1,m_2,k) \equiv \int \frac{d^dp}{(2\pi)^d} \frac{-ip^\mu}{p^2+m_1^2-i\varepsilon}\frac{-i}{(p-k)^2+m_2^2-i\varepsilon}
\quad, 
\\
&B^{\mu}_{LL}(m_1,m_2,k) \equiv \int \frac{d^dp}{(2\pi)^d} \frac{i\ p^\mu}{p^2+m_1^2+i\varepsilon}\frac{i}{(p-k)^2+m_2^2+i\varepsilon}
\quad, \\
&B^{\mu}_{RL}(m_1,m_2,k) \equiv \int \frac{d^dp}{(2\pi)^d} \frac{-i\ p^\mu}{p^2+m_1^2-i\varepsilon}\frac{i}{(p-k)^2+m_2^2+i\varepsilon}
\quad, \\
&B^{\mu}_{LR}(m_1,m_2,k) \equiv \int \frac{d^dp}{(2\pi)^d} \frac{i\ p^\mu}{p^2+m_1^2+i\varepsilon}\frac{-i}{(p-k)^2+m_2^2-i\varepsilon}
\quad.\\
\end{split}
\end{equation}
Now, we explicitly demonstrate the PV reduction for one integral from of \eqref{vectorbtype3} and we write the answer for the rest. Consider $B_{RL}$ from \eqref{vectorbtype3}. We multiply the integral expression of $B^{\mu}_{RL}(m_1,m_2,k)$ by $k^\mu$
\begin{equation}
\label{vectorbtype4}
k_\mu\,  B^\mu_{RL}(m_1,m_2,k) =(-i)(i)\int \frac{d^dp}{(2\pi)^d} \frac{\frac{1}{2}\left[(p^2+m^2)-((p-k)^2+m_2^2)+(k^2-m_1^2+m_2^2)\right]}{\left(p^2+m_1^2-i\varepsilon\right)\left((p-k)^2+m_2^2+ i\varepsilon\right)}
\quad. 
\end{equation}
Using equation \eqref{scalaratype2a} and equation  \eqref{scalaratype2b} we can write this as 
\begin{equation}
\label{vectorbtype5}
k_\mu B^\mu_{RL}(m_1,m_2,k) =\frac{i}{2}A_{L}(m_1)+\frac{i}{2} A_{R}(m_2)+\frac{k^2-m_1^2+m_2^2}{2}\ B_{RL}(m_1,m_2,k)
\quad. 
\end{equation}
Then we compare \eqref{vectorbtype1} and  \eqref{vectorbtype5} to get
\begin{equation}
\label{vectorbtype6}
B^{(1)}_{RL} = \frac{1}{2k^2}\Big[i A_{L}(m_1)+ i A_{R}(m_2)+(k^2-m_1^2+m_2^2)\ B_{RL}(m_1,m_2,k)\Big] 
\quad. 
\end{equation}
The PV formulae for the other three integrals in \eqref{vectorbtype3} are given by 
\begin{equation}
\label{vectorbtype8}
\begin{split}
B^{\mu}_{RR} &= \frac{k^\mu}{2k^2}\Big[- i A_{R}(m_2)+i A_{R}(m_1)+(k^2-m_1^2+m_2^2)\ B_{RR}(m_1,m_2,k)\Big]
\quad, \\
B^{\mu}_{LL} &= \frac{k^\mu}{2k^2}\Big[i A_{L}(m_2)-i A_{L}(m_1)+(k^2-m_1^2+m_2^2)\ B_{LL}(m_1,m_2,k)\Big]
\quad, \\
B^{\mu}_{LR} &= \frac{k^\mu}{2k^2}\Big[-i A_{L}(m_2)-i A_{R}(m_1)+(k^2-m_1^2+m_2^2)\ B_{LR}(m_1,m_2,k)\Big]
\quad. 
\end{split}
\end{equation}
Now we consider integrals with one cut propagator.  If we have only one cut propagator then one propagator is either $R$ or $L$ and the other is either $P$ or $M$. The two propagators can also be exchanged between them. So, there are eight one loop integrals with only one cut propagator
\begin{equation}
\label{vectorbtype21}
\begin{split}
&B^{\mu}_{RP}(m_1,m_2,k) \equiv \int \frac{d^dp}{(2\pi)^d} \frac{-i\ p^\mu}{p^2+m_1^2-i\varepsilon}2\pi\ \delta_+\left((p-k)^2+m_2^2\right)
\quad, \\
&B^{\mu}_{RM}(m_1,m_2,k) \equiv \int \frac{d^dp}{(2\pi)^d} \frac{-i\ p^\mu}{p^2+m_1^2-i\varepsilon}2\pi\ \delta_-\left((p-k)^2+m_2^2\right)
\quad, \\
&B^{\mu}_{PR}(m_1,m_2,k) \equiv \int \frac{d^dp}{(2\pi)^d}2\pi\ \delta_+\left(p^2+m_1^2\right) \frac{-i\ p^\mu}{(p-k)^2+m_2^2-i\varepsilon}
\quad, \\
&B^{\mu}_{MR}(m_1,m_2,k) \equiv \int \frac{d^dp}{(2\pi)^d}2\pi\ \delta_-\left(p^2+m_1^2\right) \frac{-i\ p^\mu}{(p-k)^2+m_2^2-i\varepsilon}
\quad, \\
&B^{\mu}_{LP}(m_1,m_2,k) \equiv \int \frac{d^dp}{(2\pi)^d} \frac{i\ p^\mu}{p^2+m_1^2+i\varepsilon}2\pi\ \delta_+\left((p-k)^2+m_2^2\right)
\quad, \\
&B^{\mu}_{LM}(m_1,m_2,k) \equiv \int \frac{d^dp}{(2\pi)^d} \frac{i\ p^\mu}{p^2+m_1^2+i\varepsilon}2\pi\ \delta_-\left((p-k)^2+m_2^2\right)
\quad, \\
&B^{\mu}_{PL}(m_1,m_2,k) \equiv \int \frac{d^dp}{(2\pi)^d}2\pi\ \delta_+\left(p^2+m_1^2\right) \frac{i\ p^\mu}{(p-k)^2+m_2^2+i\varepsilon}
\quad, \\
&B^{\mu}_{ML}(m_1,m_2,k) \equiv \int \frac{d^dp}{(2\pi)^d}2\pi\ \delta_-\left(p^2+m_1^2\right) \frac{i\ p^\mu}{(p-k)^2+m_2^2+i\varepsilon}
\quad. \\
\end{split}
\end{equation}
As in previous section, we present the explicit computation for only one of them and we only write the final answer for rest of them.  Consider  $B^\mu_{RP}$ from \eqref{vectorbtype21} 
and multiply it by $k^\mu$ to get 
\begin{equation}
\label{vectorbtype23}
\begin{split}
& k_\mu B^\mu_{RP}(m_1,m_2,k)\\
=& \frac{-i}{2}\int \frac{d^dp}{(2\pi)^d} \frac{\left[(p^2+m^2)-((p-k)^2+m_2^2)+(k^2-m_1^2+m_2^2)\right]}{\left(p^2+m_1^2-i\varepsilon\right)}2\pi\delta_+\left((p-k)^2+m_2^2\right)
\\
=&\frac{1}{2}\Big[-i A_{P}(m_2)+(k^2-m_1^2+m_2^2)\, B_{RP}(m_1,m_2,k)\Big]
\quad. 
\end{split}
\end{equation}
We compare this to the equation \eqref{vectorbtype2}. 
The above expression would be equal to $ k^2\, B^{(1)}_{RP}$. So the PV reduction formula for $B^\mu_{RP}$ is given by
\begin{equation}
\label{vectorbtype24}
B^{\mu}_{RP} = \frac{1}{2k^2} B^{(1)}_{RP} = \frac{k^\mu}{2k^2}\Big[-i A_{P}(m_2)+(k^2-m_1^2+m_2^2)\ B_{RP}(m_1,m_2,k)\Big]
\quad.
\end{equation} 
Following the same method, we can compute the PV formula for the rest of the integrals in \eqref{vectorbtype21}. The PV reduction formulae are given by
\begin{equation}
\label{vectorbtype25}
\begin{split}
B^{\mu}_{RM}(m_1,m_2,k) &= \frac{k^\mu}{2k^2}\left[-i A_{M}(m_2)+(k^2-m_1^2+m_2^2)\ B_{RM}(m_1,m_2,k)\right]
\quad, \\
B^{\mu}_{PR}(m_1,m_2,k) &= \frac{k^\mu}{2k^2}\left[-i A_{P}(m_1)+(k^2-m_1^2+m_2^2)\ B_{PR}(m_1,m_2,k)\right]
\quad, \\
B^{\mu}_{MR}(m_1,m_2,k) &= \frac{k^\mu}{2k^2}\left[-i A_{M}(m_1)+(k^2-m_1^2+m_2^2)\ B_{MR}(m_1,m_2,k)\right]
\quad, \\
B^{\mu}_{LP}(m_1,m_2,k) &= \frac{k^\mu}{2k^2}\left[i A_{P}(m_2)+(k^2-m_1^2+m_2^2)\ B_{LP}(m_1,m_2,k)\right]
\quad, \\
B^{\mu}_{LM}(m_1,m_2,k) &= \frac{k^\mu}{2k^2}\left[i A_{P}(m_2)+(k^2-m_1^2+m_2^2)\ B_{LM}(m_1,m_2,k)\right]
\quad, \\
B^{\mu}_{PL}(m_1,m_2,k) &= \frac{k^\mu}{2k^2}\left[i A_{P}(m_1)+(k^2-m_1^2+m_2^2)\ B_{PL}(m_1,m_2,k)\right]
\quad, \\
B^{\mu}_{ML}(m_1,m_2,k) &= \frac{k^\mu}{2k^2}\left[i A_{M}(m_1)+(k^2-m_1^2+m_2^2)\ B_{ML}(m_1,m_2,k)\right]
\quad. \end{split}
\end{equation}
Compare these expressions with \eqref{vectorbtype8}. We can see that the RHS of \eqref{vectorbtype25} has  \textit{one} $A$ type integral compared to \textit{two} $A$ type integrals in \eqref{vectorbtype8}. This is because the Dirac-delta function prohibits the presence of $A$ type integrals. This fact can be explicitly seen in \eqref{vectorbtype23}.

\subsubsection*{Integrals with two cut propagators}
\label{subsubsec:pveskbtypetwocutvector}

Now we are left with integrals with all cut propagators.  Then we have only two possibilities, either $P$ or $M$ for each of the two propagators. So there can be total four loop integrals, which are given by
\begin{equation}
\label{vectorbtype41}
\begin{split}
&B^{\mu}_{PP}(m_1,m_2,k) \equiv \int \frac{d^dp}{(2\pi)^d}2\pi\ \delta_+\left(p^2+m_1^2\right)2\pi\ \delta_+\left((p-k)^2+m_2^2\right)p^\mu
\quad, \\
&B^{\mu}_{MM}(m_1,m_2,k) \equiv \int   \frac{d^dp}{(2\pi)^d}2\pi\ \delta_-\left(p^2+m_1^2\right)2\pi\ \delta_-\left((p-k)^2+m_2^2\right)p^\mu
\quad, \\
&B^{\mu}_{PM}(m_1,m_2,k) \equiv \int  \frac{d^dp}{(2\pi)^d}2\pi\ \delta_+\left(p^2+m_1^2\right)2\pi\ \delta_-\left((p-k)^2+m_2^2\right)p^\mu
\quad, \\
&B^{\mu}_{MP}(m_1,m_2,k) \equiv \int   \frac{d^dp}{(2\pi)^d}2\pi\ \delta_-\left(p^2+m_1^2\right)2\pi\ \delta_+\left((p-k)^2+m_2^2\right)p^\mu
\quad. \\
\end{split}
\end{equation}
In equation \eqref{vectorbtype23}, we have seen that the Dirac-delta function prohibits the presence of $A$ type integrals. Now the rest of the $B^\mu$ type integrals in \eqref{vectorbtype42} have two delta functions. So following the same strategy we can show that there should not be $A$ type loop integrals in the PV formula for the rest of the $B^\mu$ type integrals. The PV formula are given by  
\begin{equation}
\label{vectorbtype42}
\begin{split}
B^\mu_{PP}(m_1,m_2,k) &= \frac{k^\mu}{2k^2}\Big[(k^2-m_1^2+m_2^2)\ B_{PP}(m_1,m_2,k)\Big]
\quad, \\
B^\mu_{MM}(m_1,m_2,k) &= \frac{k^\mu}{2k^2}\Big[(k^2-m_1^2+m_2^2)\ B_{MM}(m_1,m_2,k)\Big]
\quad, 
\\
B^\mu_{PM}(m_1,m_2,k) &= \frac{k^\mu}{2k^2}\Big[(k^2-m_1^2+m_2^2)\ B_{PM}(m_1,m_2,k)\Big]
\quad, \\
B^\mu_{MP}(m_1,m_2,k) &= \frac{k^\mu}{2k^2}\Big[(k^2-m_1^2+m_2^2)\ B_{MP}(m_1,m_2,k)\Big]
\quad.
\\
\end{split}
\end{equation}
This completes our analysis for $B$-type vector integrals. 

\subsubsection{$B^{\mu\nu}_{ab}(m_1,m_2)$ tensor integrals}
\label{subsec:pveskbtypetensor}

$B^{\mu\nu}_{ab}(m_1,m_2)$ integrals appear in loops with fermionic propagators. These tensor loop integrals can be represented as linear combination of $A$ and $B$ type scalar loop integrals. The general form of these integrals are given by the following expression 
\begin{eqnarray}
B^{\mu \nu }_{ab}(m_1,m_2;k) 
=
\mommea  
\,
(p^{\mu}p^{\nu})
\pgator_{a} (p,m_1)
\pgator_{b} ((k+p),m_{2})
\quad.
\end{eqnarray}
The $B^{\mu\nu}$ type integrals have two Lorentz indices. Then, the final answer must be a Lorentz two tensor. But there are only two available 2-tensors - $k^\mu k^\nu$ and $\eta^{\mu\nu}$. So the $B^{\mu\nu}$ should be of the form 
\begin{equation}
B^{\mu\nu}_{ab}(m_1,m_2,k) =k^\mu k^\nu B^{(21)}_{ab} +\eta^{\mu\nu} B^{(22)}_{ab}
\quad. 
\label{tensorbtype1}
\end{equation}  
$B^{(21)}_{ab}$ and $B^{(22)}_{ab}$ are constants of proportionality. We can multiply eqn \eqref{tensorbtype1} by $k^\mu k^\nu$ and $\eta^{\mu\nu}$. This gives us two equations. Solving those two equations we find expressions for $B^{(21)}_{ab}$ and $B^{(22)}_{ab}$. We again arrange sixteen  $B^{\mu\nu}_{ab}$ loop integrals by the number of cut propagators in the loops. 

\subsubsection*{Integrals with no-cut propagator \cite{Passarino:1978jh}}
\label{subsubsec:pveskbtypenocuttensor}
First consider the tensor integrals with no cut propagator. The expression for these loop integrals are given by  
\begin{equation}
\label{tensorbtype11}
\begin{split}
&B^{\mu\nu}_{RR}(m_1,m_2,k) \equiv \int \frac{d^dp}{(2\pi)^d} \frac{-ip^\mu p^\nu}{p^2+m_1^2-i\varepsilon}\frac{-i}{(p-k)^2+m_2^2-i\varepsilon}
\quad, \\
&B^{\mu\nu}_{LL}(m_1,m_2,k) \equiv \int \frac{d^dp}{(2\pi)^d} \frac{i\ p^\mu p^\nu}{p^2+m_1^2+i\varepsilon}\frac{i}{(p-k)^2+m_2^2+i\varepsilon}
\quad, \\
&B^{\mu\nu}_{RL}(m_1,m_2,k) \equiv \int \frac{d^dp}{(2\pi)^d} \frac{-i\ p^\mu p^\nu}{p^2+m_1^2-i\varepsilon}\frac{i}{(p-k)^2+m_2^2+i\varepsilon}
\quad, \\
&B^{\mu\nu}_{LR}(m_1,m_2,k) \equiv \int \frac{d^dp}{(2\pi)^d} \frac{i\ p^\mu p^\nu}{p^2+m_1^2+i\varepsilon}\frac{-i}{(p-k)^2+m_2^2-i\varepsilon}
\quad. \\
\end{split}
\end{equation}
We discussed the strategy below equation \eqref{tensorbtype1}. Now we show explicit implementation for one of the above integrals.  Consider thhe loop integral, $B^{\mu\nu}_{RL}$ from \eqref{tensorbtype11}. 
\begin{equation}
\label{tensorbtype12}
B^{\mu\nu}_{RL}(m_1,m_2,k) = \int \frac{d^dp}{(2\pi)^d} \frac{-ip^\mu p^\nu}{p^2+m_1^2-i\varepsilon}\frac{i}{(p-k)^2+m_2^2+i\varepsilon}
\quad. \\
\end{equation}
The above integral can only be a linear combination of $k^\mu k^\nu$ and $\eta^{\mu\nu}$. So we multiply equation \eqref{tensorbtype12} by $\eta^{\mu\nu}$ and $k^\mu$ we get the following two equations.
\begin{equation}
\label{tensorbtype13}
\begin{split}
k^2 B^{(21)}_{RL} + d\ B^{(22)}_{RL} = (-i)A_L(m_2) -m_1^2 B_{RL}(m_1,m_2,k)
\quad. 
\end{split}
\end{equation}
and
\begin{equation}
\label{tensorbtype14}
\begin{split}
k^2 B^{(21)}_{RL} + B^{(22)}_{RL} &= \frac{1}{2}\left[-i A_L(m_2) + (k^2-m_1^2+m_2^2)B^{(1)}_{RL}(m_1,m_2,k)\right]
\quad. 
\end{split}
\end{equation}
where $B^{(1)}_{RL}$ is defined in \eqref{vectorbtype6}. We solve eqn \eqref{tensorbtype13} and eqn \eqref{tensorbtype14}  to get the explicit expression for $B^{(21)}_{RL}$ and $B^{(22)}_{RL}$
\begin{equation}
\label{tensorbtype15}
\begin{split} 
B^{(21)}_{RL}(m_1,m_2,k) &=\frac{1}{(d-1)k^2}\Bigg[-i(d/2-1)A_L(m_2)+m_1^2B_{RL}(m_1,m_2,k)\\
&\ \ \ \ \ \ \ \ \ \ \ \ \ \ \ \ \ \ \ \ \ +(k^2-m_1^2+m_2^2)d/2\ B^{(1)}_{RL}(m_1,m_2,k)\Bigg]\qquad,\\
B^{(22)}_{RL}(m_1,m_2,k) &=\frac{1}{d-1}\Bigg[ \frac{-i}{2}A_L(m_2) -m_1^2 B_{RL}(m_1,m_2,k) -\frac{k^2-m_1^2+m_2^2}{2}B^{(1)}_{RL}\Bigg]\qquad.
\end{split}
\end{equation}
Inserting \eqref{tensorbtype15} in \eqref{tensorbtype1} we get the PV reduction for $B^{\mu\nu}_{RL}$. Following similar steps one can obtain the expressions for $B^{(21)}_{a_1 a_2}$ and $B^{(22)}_{a_1 a_2}$. They are given as follows
\begin{subequations}
\begin{eqnarray}
B^{(21)}_{RR} &=& \frac{1}{(d-1)k^2}\Big[-i(d/2-1)A_R(m_2)+m_1^2B_{RR} +(k^2-m_1^2+m_2^2)d/2\ B^{(1)}_{RR}\Big]
\quad, \label{tensorbtype18a}
\\
B^{(22)}_{RR} &=&\frac{1}{d-1}\Big[ \frac{-i}{2}A_R(m_2) -m_1^2 B_{RR} -\frac{k^2-m_1^2+m_2^2}{2}B^{(1)}_{RR}\Big]
\quad. \label{tensorbtype18b}
\end{eqnarray}
\begin{eqnarray}
B^{(21)}_{LR} &=& \frac{1}{(d-1)k^2}\Big[i(d/2-1)A_R(m_2)+m_1^2B_{LR} +(k^2-m_1^2+m_2^2)d/2\ B^{(1)}_{LR}\Big]
\quad, \label{tensorbtype18c}
\\
B^{(22)}_{LR} &=&\frac{1}{d-1}\Big[ \frac{i}{2}A_R(m_2)-m_1^2 B_{LR} -\frac{k^2-m_1^2+m_2^2}{2}B^{(1)}_{LR}\Big]
\quad. 
\label{tensorbtype18d}
\end{eqnarray}
\begin{eqnarray}
B^{(21)}_{LL} &=& \frac{1}{(d-1)k^2}\Big[i(d/2-1)A_L(m_2)+m_1^2B_{LL} +(k^2-m_1^2+m_2^2)d/2\,  B^{(1)}_{LL}\Big]
\quad, \label{tensorbtype18e}
\\
B^{(22)}_{LL} &=&\frac{1}{d-1}\Big[ \frac{i}{2}A_L(m_2)-m_1^2 B_{LL} -\frac{k^2-m_1^2+m_2^2}{2}B^{(1)}_{LL}\Big]
\quad. \label{tensorbtype18f}
\end{eqnarray}
\end{subequations}
One can see that the form of $B^{(21)}_{RR}$ \& $B^{(21)}_{RL}$ and $B^{(21)}_{LR}$ \& $B^{(21)}_{LL}$ are very similar. However, the first term in \eqref{tensorbtype18a} (and in \eqref{tensorbtype18b}) differs from the first term in \eqref{tensorbtype18c} (and in \eqref{tensorbtype18d}) by a sign. 
$B^{(21)}_{RR}$ and $B^{(22)}_{RR}$ appear unitary quantum field theory and the reduction was done in original paper \cite{Passarino:1978jh}.

\subsubsection*{Integrals with one cut propagators}
\label{subsubsec:pveskbtypeonecuttensor}

Now we want to consider tensor integrals with one cut propagator. The explicit expression for the loop integrals with one cut propagator are given by
\begin{equation}
\label{tensorbtype31}
\begin{split}
&B^{\mu\nu}_{RP}(m_1,m_2,k) \equiv \int \frac{d^dp}{(2\pi)^d} \frac{-i\ p^\mu p^\nu}{p^2+m_1^2-i\varepsilon}2\pi\ \delta_+\left((p-k)^2+m_2^2\right)
\quad, \\
&B^{\mu\nu}_{RM}(m_1,m_2,k) \equiv \int \frac{d^dp}{(2\pi)^d} \frac{-i\ p^\mu p^\nu}{p^2+m_1^2-i\varepsilon}2\pi\ \delta_-\left((p-k)^2+m_2^2\right)
\quad, \\
&B^{\mu\nu}_{PR}(m_1,m_2,k) \equiv \int \frac{d^dp}{(2\pi)^d}2\pi\ \delta_+\left(p^2+m_1^2\right) \frac{-i\ p^\mu p^\nu}{(p-k)^2+m_2^2-i\varepsilon}
\quad, \\
&B^{\mu\nu}_{MR}(m_1,m_2,k) \equiv \int \frac{d^dp}{(2\pi)^d}2\pi\ \delta_-\left(p^2+m_1^2\right) \frac{-i\ p^\mu p^\nu}{(p-k)^2+m_2^2-i\varepsilon}
\quad, \\
&B^{\mu\nu}_{LP}(m_1,m_2,k) \equiv \int \frac{d^dp}{(2\pi)^d} \frac{i\ p^\mu p^\nu}{p^2+m_1^2+i\varepsilon}2\pi\ \delta_+\left((p-k)^2+m_2^2\right)
\quad, \\
&B^{\mu\nu}_{LM}(m_1,m_2,k) \equiv \int \frac{d^dp}{(2\pi)^d} \frac{i\ p^\mu p^\nu}{p^2+m_1^2+i\varepsilon}2\pi\ \delta_-\left((p-k)^2+m_2^2\right)
\quad, \\
&B^{\mu\nu}_{PL}(m_1,m_2,k) \equiv \int \frac{d^dp}{(2\pi)^d}2\pi\ \delta_+\left(p^2+m_1^2\right) \frac{i\ p^\mu p^\nu}{(p-k)^2+m_2^2+i\varepsilon}
\quad, \\
&B^{\mu\nu}_{ML}(m_1,m_2,k) \equiv \int \frac{d^dp}{(2\pi)^d}2\pi\ \delta_-\left(p^2+m_1^2\right) \frac{i\ p^\mu p^\nu}{(p-k)^2+m_2^2+i\varepsilon}
\quad. \\
\end{split}
\end{equation}
Following the steps shown in last section, It is straightforward to do the PV reduction for all loop integrals in \eqref{tensorbtype31}. In fact these can be written in a single equation as the following:
\begin{equation}
\label{tensorbtype37}
\begin{split}
B^{(21)}_{a_1a_2}(m_1,m_2,k) &=\frac{1}{(d-1)k^2}\left[m_1^2 B_{a_1a_2} +(k^2-m_1^2+m_2^2)d/2\ B^{(1)}_{a_1a_2} \right],\\
B^{(22)}_{a_1a_2}(m_1,m_2,k) &=\frac{1}{(d-1)}\left[-m_1^2 B_{a_1a_2} -(k^2-m_1^2+m_2^2)/2\ B^{(1)}_{a_1a_2} \right]\quad.\\
\end{split}
\end{equation}
where $a_1$ can be $P$ or $M$ and $a_2$ can be $R$ or $L$. 

\subsubsection*{Integrals with two cut propagators}
\label{subsubsec:pveskbtypetwocuttensor}

There are four loop integrals with two cuts. They are given by
\begin{equation}
\label{tensorbtype51}
\begin{split}
&B^{\mu\nu}_{PP}(m_1,m_2,k) \equiv \int \frac{d^dp}{(2\pi)^d}2\pi\ \delta_+\left(p^2+m_1^2\right)2\pi\ \delta_+\left((p-k)^2+m_2^2\right) p^\mu p^\nu
\quad, \\
&B^{\mu\nu}_{MM}(m_1,m_2,k) \equiv \int  \frac{d^dp}{(2\pi)^d}2\pi\ \delta_-\left(p^2+m_1^2\right)2\pi\ \delta_-\left((p-k)^2+m_2^2\right)p^\mu p^\nu
\quad, \\
&B^{\mu\nu}_{PM}(m_1,m_2,k) \equiv \int  \frac{d^dp}{(2\pi)^d}2\pi\ \delta_+\left(p^2+m_1^2\right)2\pi\ \delta_-\left((p-k)^2+m_2^2\right)p^\mu p^\nu
\quad, \\
&B^{\mu\nu}_{MP}(m_1,m_2,k) \equiv \int   \frac{d^dp}{(2\pi)^d}2\pi\ \delta_-\left(p^2+m_1^2\right)2\pi\ \delta_+\left((p-k)^2+m_2^2\right)p^\mu p^\nu
\quad. \\
\end{split}
\end{equation}
The PV reduction of these integrals give the following  expressions for $B^{(21)}_{a_3a_4}$ and  $B^{(22)}_{a_3a_4}$ 
\begin{equation}
\begin{split}
B^{(21)}_{a_3a_4}(m_1,m_2,k) &=\frac{1}{(d-1)k^2}\left[m_1^2\,  B_{a_3a_4} +(k^2-m_1^2+m_2^2)d/2\ B^{(1)}_{a_3a_4} \right]
\quad, \\
B^{(22)}_{a_3a_4}(m_1,m_2,k) &=\frac{1}{(d-1)}\left[-m_1^2\,  B_{a_3a_4} -(k^2-m_1^2+m_2^2)/2\ B^{(1)}_{a_3a_4} \right]
\quad. \\
\end{split}
\label{tensorbtype53}
\end{equation}
here $a_3$ and $a_4$ are either $P$ or $M$.

\section{Open Yukawa theory}
\label{sec:bjropenyukawa}

In this section, we study the open-Yukawa theory; we write down the most general action of a  open QFT with one fermion and one real scalar. Then we compute the scalar tadpole and the self-energy correction to the fermion and the scalar field. The loop diagrams in this theory are evaluated using the PV reduction described in the previous section. The explicit expression for the $B$ type loop integrals that are used in this section can be found in the appendix \ref{sec:fermionicpv} . 

First, we discuss that the scalar tadpole which can be removed by introducing a local counterterm. Then we show that the divergence in the correction to the scalar propagator can also be removed by a local counterterm \footnote{However, there are other theories where the correction to the scalar propagator is non-local}. But the correction to the fermion propagator has a \emph{non-local divergence}, thus cannot be removed by local counter-terms. These divergence structures are evident from the PV reduction formula, which is absent in the unitary Yukawa theory (see the discussion in the appendix \ref{sec:fermionicpv}).

\subsection{Action for the open Yukawa theory}
\label{subsec:bjropenyukawaaction}

The action for unitary Yukawa theory \footnote{This is not the most general renormalizable unitary Yukawa theory in $D=4-\epsilon$. One can also consider Yukawa term involving $\gamma_5$. However the extension of this theory is enough to convey our point. We also considered open yukawa theory with $\gamma_5$ vertices. The essential feature of non-local divergence remains. We do not present the computation of that theory to reduce volume of the paper without compromising with the essential point.}, 
\begin{equation}
\begin{split}
S[\phi, \psi] =&\int d^4x\Bigg[-\frac{1}{2}Z_{\phi}(\partial \phi)^2 - \frac{1}{2}m_{\phi}^2 \phi^2 + iZ_{\psi}\,  \bar{\psi} \slashed{\partial} \psi -m_{\psi}  \bar{\psi}\psi 
 \\
&- \left(\frac{\lambda_3}{3!}\phi^3+\frac{\lambda_4}{4!} \phi^4 + y \phi \bar{\psi}\psi \right) \Bigg]\,.
\end{split}
\end{equation}
The SK action for the open Yukawa theory can be constructed from the above unitary action through the following steps. First we double the degrees of freedom,
\begin{eqnarray}
\bigl(\phi,\psi\bigr) \longrightarrow     \bigl(\phi_R,\psi_R\bigr),\, \bigl(\phi_L,\,\psi_L\bigr)\,.
\end{eqnarray}
$R$ ($L$) fields evolve along the forward (backward) time respectively in the SK contour. The backward time evolution results in an opposite sign in the action. The SK action is given by
\begin{align*}
S_{SK}^{\text{unitary}} = S[\phi_R, \psi_R] - S[\phi_L, \psi_L]\,.
\end{align*}
The above action has no mixed $R$, $L$ term and the couplings are real; thus it is unitary. To make the action open, one needs to complexify the existing couplings and add all possible Feynman-Vernon influence phases \cite{Feynman:1963fq}. The Feynman-Vernon influence phases are $R$, $L$ interaction terms with complex couplings. Finally, the action for open Yukawa theory is given by,
\begin{equation}\label{general open action of Yukawa}
S_{SK}= \int d^4x \Big( \mathcal{L}_{\phi} +\mathcal{L}_{\psi} +\mathcal{L}_{\text{Yuk}}\Big)\,,
\end{equation}
where $\mathcal{L}_\phi$ is the action for the open scalar field theory \citep{Avinash:2017asn}.
 \begin{equation}
\begin{split}
\mathcal{L}_\phi &
=-\Bigl[ \frac{1}{2} z_\phi\ (\partial \phir)^2 + \frac{1}{2} m_\phi^2  \phir^2+ \frac{\lambda_3}{3!}\phir^3+\frac{\lambda_4}{4!} \phir^4+\frac{\sigma_3}{2!}\phir^2\phil + \frac{\sigma_4}{3!} \phir^3 \phil \Bigr]\\
&\qquad+\Bigl[ \frac{1}{2} z_\phi^\ast (\partial \phil)^2 + \frac{1}{2} {m_\phi^2}^\ast \phil^2 +\frac{\lambda_3^\ast }{3!}\phil^3+ \frac{\lambda_4^\ast}{4!} \phil^4+ \frac{\sigma_3^\ast }{2!}\phil^2\phir+\frac{\sigma_4^\ast}{3!} \phil^3 \phir\Bigr] \\
&\qquad + i  \Bigl[ z_\Delta\ (\partial \phir)\cdot (\partial \phil)  + m^2_{\phi\Delta} \phir \phil+\frac{{\lambda_\Delta}}{2!2!} \phir^2 \phil^2 \Bigr]\,.
\end{split}
\end{equation} 
where $\sigma_3,\,\sigma_4,\,\lambda_{\Delta}$ are the couplings corresponding to the $R\,,L$ mixing terms in the action.\\
$\mathcal{L}_{\psi}$ in \eqref{general open action of Yukawa} is the open Dirac action with a mass term.
\begin{equation}
\begin{split}
\mathcal{L}_\psi &=
- \Bigl[ z_{\psi} \psibr (-i \slashed{\partial}) \psir + m_{\psi} \psibr  \psir  \Bigr] 
+  \Bigl[ z_{\psi}^\ast \psibl (-i \slashed{\partial}) \psil + m_{\psi}^\ast\psibl \psil \Bigr]\\
& + i  \Bigl[ z_{\psi\Delta}\  \psibr (-i \slashed{\partial}) \psil   +  m_{\psi\Delta} \psibr\psil \ \Bigr] 
+ i  \Bigl[ \hat{z}_{\psi\Delta}\  \psibl (-i \slashed{\partial}) \psir   + \hat{m}_{\psi\Delta} \psibl   \psir \ \Bigr]\,.
\end{split}
\end{equation}
$\mathcal{L}_{\text{Yuk}}$ include all possible open Yukawa interaction terms. 
\begin{equation}
\begin{split}
\mathcal{L}_{\text{Yuk}} =&
- \Bigl[y\phir \psibr  \psir +y_{\sigma}  \phil \psibr  \psir \Bigr] 
- \Bigl[ y_{\kappa}\phi_R \bar{\psi}_R \psi_L + y_\rho \phi_R \bar{\psi}_L \psi_R \Bigr]\\
&+ \Bigl[ y^\ast \phil \psibl   \psil + y_{\sigma}^\ast \phir \psibl \psil \Bigr]  
+ \Bigl[   y_{\kappa}^\ast \phi_L \bar{\psi}_R  \psi_L + y_\rho^* \phi_L \bar{\psi}_L   \psi_R \Bigr] \,,\\
\end{split}
\end{equation}
where $y_\sigma,\,y_\kappa,\,y_\rho$ are $R$, $L$ mixing terms in the above Yukawa interaction terms.

A point to notice that we have chosen the coupling constants in the $L$ branch are complex conjugate to that of the $R$ branch. The couplings corresponding to $L,\,R$ mixing terms are chosen to be such that the action respects the SK CPT symmetry \cite{Sieberer:2015svu}. In order to determine the Lindblad conditions (the trace-preserving conditions) one can write the action in the Lindblad form \cite{Sieberer:2015svu} and one finds that the all of the couplings are not independent; they follow certain constraints. We call these \emph{Lindblad conditions}. In open Yukawa theory, the Lindblad conditions are the following. 
\begin{subequations}\label{Lindblad_conditions}
\begin{eqnarray}
&&\im\, m_\phi^2=    m^2_{\phi \Delta}\,,
\qquad\qquad\qquad
2\im\, m_\psi=    m_{\psi \Delta} + \hat{m}_{\psi \Delta}\,,
\\
&&
\im\, z_\phi =z_{\phi \Delta}\,,
\qquad\qquad\qquad
2\im\,  z_\psi =z_{\psi \Delta} + \hat{z}_{\psi \Delta}\,,\\
&&\im \lambda_{3}+3\, \im \lambda_{3\sigma}=0\,,
\qquad\qquad
\im \lambda_{4}+4\im \lambda_{4\sigma}=3\lambda_{\phi\Delta}\,,
\\
&&
\im\, y+ \im\, y_\sigma+\im\, y_\kappa+\im\, y_\rho = 0 \,.
\end{eqnarray}
\end{subequations}
One can check that these conditions ensure that the SK action vanishes if we set $\phir=\phil$ and $\psir=\psil$.

\subsection{Propagators}
\label{subsec:bjropenpropagator}

There are four SK propagators in the $R$-$L$ basis for both scalar and the fermionic fields. The explicit derivation of the propagator using $i\varepsilon$ prescription for scalar field theory can be found in \cite{Avinash:2017asn}. The fermionic propagators are straightforward to write from the scalar propagators. The propagators and their diagrammatic representations are shown in fig. \ref{fig:scalarpropagator} and fig. \ref{fig:diracpropagator}.
\begin{figure}[ht]
\begin{center}
    
\begin{tikzpicture} 
\phipropagatorr{0}{0}{2}{0}{p\rightarrow}
\node at (-2,0) {R :};
\node at (-1,0) {$\phir$ };
\node at (3,0) {$\phir$};
\node at (7,0) {$\frac{-i}{p^2+m_\phi^2-i\varepsilon}$};

\phipropagatorp{0}{-1}{2}{-1}{p\rightarrow}
\node at (-2,-1) {P :};
\node at (-1,-1) {$\phir$ };
\node at (3,-1) {$\phil$};
\node at (7,-1) {$2\pi \delta_{+}(p^2+m_\phi^2)
$};

\phipropagatorm{0}{-2}{2}{-2}{p\rightarrow}
\node at (-2,-2) {M :};
\node at (-1,-2) {$\phil$ };
\node at (3,-2) {$\phir$};
\node at (7,-2) {$2\pi \delta_{-}(p^2+m_\phi^2)
$};

\phipropagatorl{0}{-3}{2}{-3}{p\rightarrow}
\node at (-2,-3) {L :};
\node at (-1,-3) {$\phil$ };
\node at (3,-3) {$\phil$};
\node at (7,-3) {$\frac{i}{p^2+m_\phi^2+i\varepsilon}$};
\end{tikzpicture}

\end{center}

\caption{SK propagators for the scalar field}
\label{fig:scalarpropagator}
\end{figure}

\begin{figure}[ht]
\begin{center}

\begin{tikzpicture} 

\psipropagatorr{0}{0}{2}{0}{ $p\rightarrow$}
\node at (-2,0) {$R^f$ :};
\node at (-1,0) {$\psibr$ };
\node at (3,0) {$\psir$};

\node at (7,0) { $\frac{-i(-\slashed{p}+m_\psi)}{p^2+m_\psi^2-i\varepsilon}$};

\psipropagatorp{0}{-1}{2}{-1}{ $p\rightarrow$}
\node at (-2,-1) {$P^f $:};
\node at (-1,-1) {$\psibr$  };
\node at (3,-1) {$\psil$ };

\node at (7,-1) { $ (-\slashed{p}+m_\psi)\ 2\pi \delta_+( p^2+m_\psi^2)$};

\psipropagatorm{0}{-2}{2}{-2}{ $p\rightarrow$}
\node at (-2,-2) {$M^f$ :};
\node at (-1,-2) {$\psibl$ };
\node at (3,-2) {$\psir$ };

\node at (7,-2) { $ (-\slashed{p}+m_\psi)\ 2\pi \delta_-( p^2+m_\psi^2)$};

\psipropagatorl{0}{-3}{2}{-3}{ $p\rightarrow$}
\node at (-2,-3) {$L^f$ :};
\node at (-1,-3) {$\psibl$ };
\node at (3,-3) {$\psil$ };

\node at (7,-3) { $\frac{i(-\slashed{p}+m_\psi)}{p^2+m_\psi^2+i\varepsilon}$};

\end{tikzpicture} 

\end{center}
\caption{SK propagators for the Dirac field}
\label{fig:diracpropagator}
\end{figure} 

\subsection{Feynman rules}

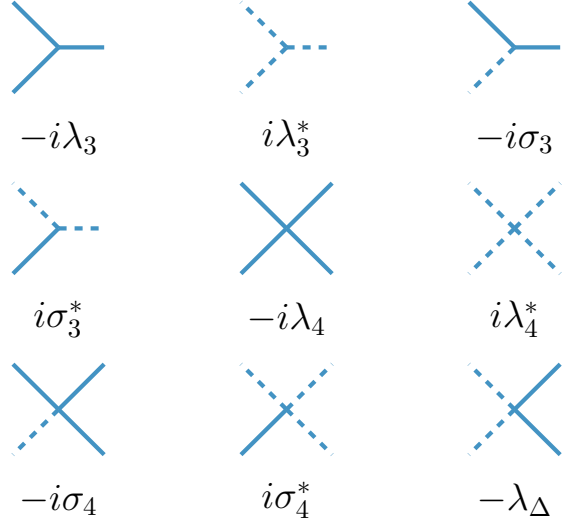
\begin{figure}[ht] \label{Spin propagator}
\begin{center}
\begin{tikzpicture}[line width=1 pt, scale=0.6]

\begin{scope}[shift={(0,4)}]
\draw [phir, ultra thick]  (0,0) -- (1,0);
\draw [phir, ultra thick]  (0,0) -- (-1,1);
\draw [phir, ultra thick]  (0,0) -- (-1,-1);
\node at (0,-2) {\Large $-i\lambda_3$};    
\end{scope} 

\begin{scope}[shift={(5,4)}]
\draw [phil, ultra thick]  (0,0) -- (1,0);
\draw [phil, ultra thick]  (0,0) -- (-1,1);
\draw [phil, ultra thick]  (0,0) -- (-1,-1);
\node at (0,-2) {\Large $i\lambda_3^\ast $};    
\end{scope} 

\begin{scope}[shift={(10,4)}]
\draw [phir, ultra thick]  (0,0) -- (1,0);
\draw [phir, ultra thick]  (0,0) -- (-1,1);
\draw [phil, ultra thick]  (0,0) -- (-1,-1);
\node at (0,-2) {\Large $-i\sigma_3$};    
\end{scope}

\begin{scope}[shift={(0,0)}]
\draw [phil, ultra thick]  (0,0) -- (1,0);
\draw [phil, ultra thick]  (0,0) -- (-1,1);
\draw [phir, ultra thick]  (0,0) -- (-1,-1);
\node at (0,-2) {\Large $i\sigma_3^\ast$};    
\end{scope} 

\begin{scope}[shift={(5,0)}]
\draw [phir, ultra thick]  (0,0) -- (1,1);
\draw [phir, ultra thick]  (0,0) -- (1,-1);
\draw [phir, ultra thick]  (0,0) -- (-1,1);
\draw [phir, ultra thick]  (0,0) -- (-1,-1);
\node at (0,-2) {\Large $-i\lambda_4$};    
\end{scope} 

\begin{scope}[shift={(10,0)}]
\draw [phil, ultra thick]  (0,0) -- (1,1);
\draw [phil, ultra thick]  (0,0) -- (1,-1);
\draw [phil, ultra thick]  (0,0) -- (-1,1);
\draw [phil, ultra thick]  (0,0) -- (-1,-1);
\node at (0,-2) {\Large $i\lambda_4^\ast $};    
\end{scope}

\begin{scope}[shift={(0,-4)}]
\draw [phir, ultra thick]  (0,0) -- (1,1);
\draw [phir, ultra thick]  (0,0) -- (1,-1);
\draw [phir, ultra thick]  (0,0) -- (-1,1);
\draw [phil, ultra thick]  (0,0) -- (-1,-1);
\node at (0,-2) {\Large $-i\sigma_4$};    
\end{scope}

\begin{scope}[shift={(5,-4)}] 
\draw [phil, ultra thick]  (0,0) -- (1,1);
\draw [phil, ultra thick]  (0,0) -- (1,-1);
\draw [phil, ultra thick]  (0,0) -- (-1,1);
\draw [phir, ultra thick]  (0,0) -- (-1,-1);
\node at (0,-2) {\Large $i\sigma_4^\ast $};    
\end{scope}

\begin{scope}[shift={(10,-4)}]
\draw [phir, ultra thick]  (0,0) -- (1,1);
\draw [phir, ultra thick]  (0,0) -- (1,-1);
\draw [phil, ultra thick]  (0,0) -- (-1,1);
\draw [phil, ultra thick]  (0,0) -- (-1,-1);
\node at (0,-2) {\Large $-\lambda_\Delta$};    
\end{scope}

\end{tikzpicture}
\end{center}
\caption{Feynman rules for scalar couplings}
\label{fig:macrotree1} 
\end{figure}

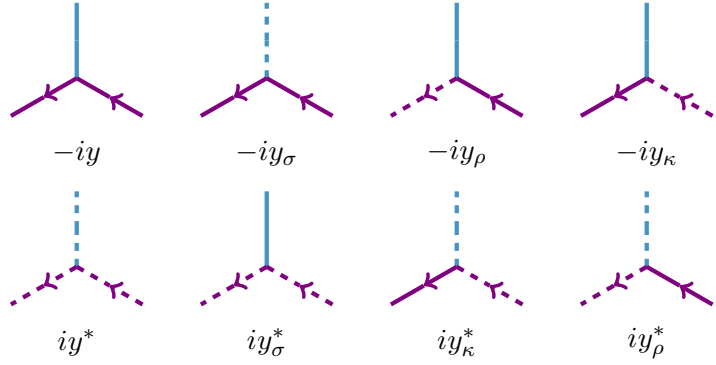
\begin{figure}[h] 
\begin{center} 
\begin{tikzpicture}[line width=1 pt, scale=0.5]

\begin{scope}[shift={(0,0)}]

\phipropagatorr{0}{0}{0}{2}{} 
\begin{scope}[rotate=120]
\psipropagatorr{0}{0}{0}{2}{} 
\end{scope} 
\begin{scope}[rotate=240]
\psibarpropagatorr{0}{0}{0}{2}{} 
\end{scope} 

\node at (0,-2) { $-iy $};
    
\end{scope}

\begin{scope}[shift={(5,0)}]

\phipropagatorl{0}{0}{0}{2}{} 
\begin{scope}[rotate=120]
\psipropagatorr{0}{0}{0}{2}{} 
\end{scope} 
\begin{scope}[rotate=240]
\psibarpropagatorr{0}{0}{0}{2}{} 
\end{scope} 

\node at (0,-2) {$-iy_\sigma $};
    
\end{scope}

\begin{scope}[shift={(10,0)}]

\phipropagatorr{0}{0}{0}{2}{} 
\begin{scope}[rotate=120]
\psipropagatorl{0}{0}{0}{2}{} 
\end{scope} 
\begin{scope}[rotate=240]
\psibarpropagatorr{0}{0}{0}{2}{} 
\end{scope} 

\node at (0,-2) { $-iy_\rho$};
    
\end{scope}

\begin{scope}[shift={(15,0)}]

\phipropagatorr{0}{0}{0}{2}{} 
\begin{scope}[rotate=120]
\psipropagatorr{0}{0}{0}{2}{} 
\end{scope} 
\begin{scope}[rotate=240]
\psibarpropagatorl{0}{0}{0}{2}{} 
\end{scope} 

\node at (0,-2) { $-iy_\kappa$};
    
\end{scope}


\begin{scope}[shift={(0,-5)}]

\phipropagatorl{0}{0}{0}{2}{} 
\begin{scope}[rotate=120]
\psipropagatorl{0}{0}{0}{2}{} 
\end{scope} 
\begin{scope}[rotate=240]
\psibarpropagatorl{0}{0}{0}{2}{} 
\end{scope} 

\node at (0,-2) { $iy^\ast$};
    
\end{scope}

\begin{scope}[shift={(5,-5)}]

\phipropagatorr{0}{0}{0}{2}{} 
\begin{scope}[rotate=120]
\psipropagatorl{0}{0}{0}{2}{} 
\end{scope} 
\begin{scope}[rotate=240]
\psibarpropagatorl{0}{0}{0}{2}{} 
\end{scope} 

\node at (0,-2) {$iy_\sigma^\ast$};
    
\end{scope}

\begin{scope}[shift={(10,-5)}]

\phipropagatorl{0}{0}{0}{2}{} 
\begin{scope}[rotate=120]
\psipropagatorr{0}{0}{0}{2}{} 
\end{scope} 
\begin{scope}[rotate=240]
\psibarpropagatorl{0}{0}{0}{2}{} 
\end{scope} 

\node at (0,-2) { $iy_\kappa^\ast$};

\end{scope}

\begin{scope}[shift={(15,-5)}]

\phipropagatorl{0}{0}{0}{2}{} 
\begin{scope}[rotate=120]
\psipropagatorl{0}{0}{0}{2}{} 
\end{scope} 
\begin{scope}[rotate=240]
\psibarpropagatorr{0}{0}{0}{2}{} 
\end{scope} 

\node at (0,-2) { $iy_\rho^\ast$};
    
\end{scope}
 
\end{tikzpicture} 
\end{center}
\caption{Feynman rules for scalar fermion couplings}
\label{fig3pt} 
\end{figure}

The Feynman rules for the cubic and quartic vertices are shown in fig. \ref{fig:macrotree1} and in fig. \ref{fig3pt}.

\begin{figure}
\begin{center}
\begin{tikzpicture}[line width=1 pt, scale=0.7] 

\begin{scope}[shift={(0,0)}]
\draw[ultra thick,violet] (-1,0) arc (180:360:1);
\draw[ultra thick,blue] (1,0) arc (0:90:1);
\draw[ultra thick,blue,dashed] (0,1) arc (90:180:1);

\node at (-1.5,0) {$1$};
\node at (1.5,0) {$2$};
\node at (1,0) {$\times$};
\node at (-1,0) {$\times$};

\node at (0,-2) {$B^{f}_{RP}$};

\end{scope}

%
%
%
%

\end{tikzpicture}
\end{center}
\caption{One example of fermionic loop diagram in SK theory }
\label{fig:BfRP}
\end{figure}
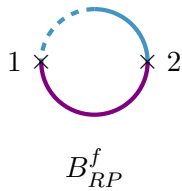

Given the Feynman rules for vertices and the propagators, one can compute the correction to the self energies for the scalar and the fermion. Since we have already fixed a nomenclature for the scalar loop integrals, we do the same for the fermionic loop integrals. The fermionic nature of an integral are mentioned in the superscript. For example, let us consider a Feynman diagram with one scalar and one fermionic internal propagator as shown in fig.  \ref{fig:BfRP}. We call this diagram $B^f_{RP}$. We can always choose the first propagator (here the $R$ propagator) either to be fermionic or to be bosonic. But, to avoid any ambiguity, we choose the first propagator (from the left) to be fermionic in $B$-type loop integrals. 

\subsection{Scalar tadpoles}
\label{subsec:bjropenyukawascalartadpole}

It is well known from the unitary Yukawa theory that there are tadpole diagrams which contribute to the one point function of the scalar field. The tadpole of a fermionic field vanishes due to underlying Lorentz invariance of the QFT. The contribution from these diagrams can be removed from the theory by introducing counter terms in the action. In open Yukawa theory there are two types of tadpole diagrams: contribution from scalar and the contribution from fermion. The Feynman diagrams can be found in appendix \ref{subsec:scalartadpolediagram} (fig \ref{fig:philtadpole1}). The contribution from all of these diagrams are the following.
\begin{equation}
\begin{split}
&\frac{-i\lambda_3}{2}A_{R}
+ \frac{i\lambda_{3\sigma}^\star}{2}A_{L}
+(-i\lambda_{3\sigma}) A_{P}\\ 
& + (-iy) A^f_{R} + (iy_{\sigma}^\ast) A^f_{L}
+ (-iy_\kappa) A^f_{P} + (-iy_\rho)A^f_{M}\\
=&\frac{-i}{(4\pi)^2}\frac{1}{d-4}\ \Big[(m_\phi^2)(\lambda_3-\sigma_3^\star +2\sigma_3)-(8 m_\psi^3) (y - y_{\sigma}^\ast + y_\kappa + y_\rho )\Big] + ... \,\,.
\end{split}
\end{equation}
The above divergent piece is local. The scalar tadpole can be removed by a counter-term of the form $\kappa\phir$ to the action where $\kappa$ is given by 
\begin{equation}
\begin{split}
\frac{i}{(4\pi)^2}\frac{1}{d-4}\ \Big[(m_\phi^2)(\lambda_3-\sigma_3^\star +2\sigma_3)-(8 m_\psi^3) (y - y_{\sigma}^\ast + y_\kappa + y_\rho )\Big]   \,\,.
\end{split}
\end{equation}

\subsection{Mass renormalization of the scalar field}
\label{subsec:bjropenyukawascalarmass}

In the action, there are three scalar quadratic terms. Here we compute the one-loop correction to $m^2$. It receives a contribution from scalar tadpole diagrams due to scalar quadratic interactions and from thhe scalar \& fermionic bubble diagrams due to the scalar cubic couplings and Yukawa couplings. The relevant Feynman diagrams can be found in the appendix \ref{subsec:scalarmassdiagram}. 
 
The self energy of the scalar field in dimensional regularization is given by
\begin{equation}
\begin{split}
&-iz_\phi \ k^2 -i {m_\phi}^2 \\
&-\frac{i}{(4\pi)^2}\Big[\frac{1}{d-4}+\frac{1}{2}(\gamma_E-1-\ln(4\pi))\Big] \left[\lambda_4  -i \lambda_\Delta +2\sigma_4 \right]\left(\re\,m^2\right)\\
&-\frac{i}{(4\pi)^2}\Big[\frac{1}{d-4}+\frac{1}{2}(\gamma_E-1-\ln(4\pi))\Big] \Bigl[ (\lambda_3)^2 -(\sigma_3^\ast)^2+
2\left\{\lambda_3\sigma_3+|\sigma_3|^2 \right\}\ \Bigr] \\
&+ \frac{4i}{(4\pi)^2} \Big[\frac{1}{d-4}+\frac{1}{2}(\gamma_E-1-\ln(4\pi))\Big] \Bigl[(k^2+6m^2_\psi) (y+y_\sigma^\ast)(y-y_\sigma^\ast+y_{\kappa}+y_\rho)\Bigr]\,.
\end{split}
\end{equation}
The second, third, fourth line in the above expression comes from the scalar $A$-type, scalar $B$-type and fermionic $B$-type integrals respectively. We can see that the divergence is local and thus can be removed by introducing a local counter-term to the action. One can compute the scale dependence of the mass of the scalar field. The anomalous dimension of the scalar field is given by 
\begin{equation}
\begin{split}
\gamma_\phi \equiv \frac{1}{2}\frac{d\ \ln\,  z_\phi}{d\ \ln\,  \mu} = -\frac{2}{(4\pi)^2}\,(y+y_{\sigma}^\star)(y-y_{\sigma}^\star + y_{\kappa} + y_{\rho})\,.
\end{split}
\end{equation}
If we set $y_{\sigma}=y_{\kappa}=y_{\rho}=0$, we get back the result of unitary theory. The beta function for the mass of the scalar field is given by,
\begin{equation}
\begin{split}
\frac{d}{d\ \ln\,  \mu} \left(m_\phi^2\right)  =& \frac{4\, m_\phi^2-24\, m_\psi^2}{(4\pi)^2}(y+y_\sigma^\star)(y-y_\sigma^\star + y_\rho +y_\kappa)\\
&+\frac{1}{(4\pi)^2}\Bigl[ (\lambda_3)^2 -(\sigma_3^\star)^2+
2\left\{\lambda_3\sigma_3+|\sigma_3|^2 \right\}\ \Bigr] \\
&+\frac{1}{(4\pi)^2} \left[\lambda_4  -i \lambda_\Delta +2\sigma_4 \right]\left(\re\,m^2_\phi\right) \,.
\end{split}
\end{equation}

\subsection{Mass renormalization of fermionic field}
\label{subsec:bjropenyukawafermionmass}

In this section, we compute  the correction to the fermionic propagator  and show that it has a very different structure from that of the scalar propagator. There is a non-local divergence in the correction to the fermionic propagator.

The Feynman diagrams which represents that one loop correction to the self energy correction to the fermionic propagators can be found in the appendix \ref{subsec:fermionicmassdiagram}. $B$-type diagrams with one fermionic and one scalar propagator contribute to the fermionic self energy correction. The one loop correction to the fermionic propagator consists of the following contributions
\begin{equation}
\label{fermion prop correction}
\begin{split}
& (-iy)^2 B^f_{RR}(k,m_\psi,m_\phi) + (iy_\rho^\star)(iy_\kappa^\star) B^f_{LL}(k,m_\psi,m_\phi)\\
& + (-iy_\sigma)^2 B^f_{RL}(k,m_\psi,m_\phi) + (-iy_\rho)(-iy_\kappa)B^f_{LR}(k,m_\psi,m_\phi)\\
& + (-iy)(iy_\kappa^\star)B^f_{PM}(k,m_\psi,m_\phi) + (iy_\rho^\star)(-iy)B^f_{MP}(k,m_\psi,m_\phi) \\
& + (-iy_\rho)(-iy_\kappa)B^f_{PP}(k,m_\psi,m_\phi) + (-iy_\rho)(-iy_\sigma)B^f_{MM}(k,m_\psi,m_\phi)\\
& + (-iy_\sigma)(-iy) B^f_{RP}(k,m_\psi,m_\phi) + (-iy)(-iy_\kappa)B^f_{PR}(k,m_\psi,m_\phi)\\
& + (-iy)(-iy_\sigma)B^f_{RM}(k,m_\psi,m_\phi) + (-iy_\rho)(-iy)B^f_{MR}(k,m_\psi,m_\phi)\\
& + (iy_\rho^\ast)(-iy_\kappa)B^f_{LP}(k,m_\psi,m_\phi) + (-iy_\sigma)(iy_\kappa^\ast)B^f_{PL}(k,m_\psi,m_\phi)\\
& + (-iy_\rho)(iy_\kappa^\ast) B^f_{LM}(k,m_\psi,m_\phi) + (iy_\rho^\ast)(-iy_\sigma)B^f_{ML}(k,m_\psi,m_\phi)\,.
\end{split}
\end{equation} 
To compute the loop integrals in \eqref{fermion prop correction}, we use the PV reduction formula which is  done in the appendix \ref{sec:fermionicpv}. If we substitute the PV reduction for the fermionic integrals,  we find that the divergence piece of the one loop contribution is non-local in the external momentum. To see that explicitly, let us choose an integral from the above set of integrals, say $B^f_{PR}(k,m_\psi,m_\phi)$. This integral was chosen in particular, because the combination of coupling constant that appears with this integral does not appear with any other integral (and hence there is no scope of cancellation). The PV reduction of this integral is given by
\begin{eqnarray}
\label{eq:pvBPR}
B^{f}_{PR}(k,m_{\psi},m_{\phi}) &=& m_{\psi} B_{PR}\left(k,m_{\psi},m_{\phi}\right) -\slashed{k}\ \frac{\ i \ A_P\left(m_{\psi}\right)-(-k^2+m_\psi^2-m_\phi^2) B_{PR}\left(k,m_{\psi},m_{\phi}\right)}{2 k^2} \,.
\nonumber\\
\end{eqnarray}  
$A_P$ and $B_{PR}$ are discussed in the section \S\ref{sec:pveskatype} and in the section \ref{subsec:pveskbtypescalar} respectively. The divergence structure of $A_P$ and $B_{PR}$ are the following.
\begin{eqnarray}
&& A_P(m_\psi) \sim \frac{m_\psi^2}{(4\pi)^2}\frac{2}{d-4}\,, \nonumber\\
&& B_{PR}(k,m_\psi,m_\phi) \sim \frac{i}{(4\pi)^2}\frac{k^2-m_\psi^2+m_\phi^2}{2k^2} \frac{2}{d-4} \,.
\end{eqnarray}
Substituting these two back in \eqref{eq:pvBPR} we get,
\begin{align*}
B^{f}_{PR}(k,m_{\psi},m_{\phi}) \sim  \frac{i}{(4\pi)^2}\left[ m_\psi(k^2-m_\psi^2+m_\phi^2) -\slashed{k}\Big(m_\psi^2 + \frac{(k^2-m_\psi^2+m_\phi^2)^2}{2k^2}\Big) \right] \nonumber\\
\times \frac{1}{2k^2}\frac{2}{d-4}\,.
\end{align*}
This is clearly non-local since it has $k^2$ in the denominator 
. This divergence can not be removed with a local counter-term. One can notice that the divergence does not go away in the limit, $m_\phi^2 = m_\psi^2$ \footnote{A detailed discussion is done in the appendix \ref{sec:fermionicpv}.} unlike the $B$ type scalar integrals. This is a distinct feature of the fermionic $B$ type integrals. We do not have any clear understanding of any of this divergences.

Let us now evaluate the self energy correction to the fermion and show that the non-local divergences does not disappear in the self energy correction. Adding the contribution from all diagrams given in the fig. \ref{fig:fermionselfenergy}, we get the correction to be non-local and UV divergent. This feature persists in the equal mass limit of the scalar and the fermion. For sake of brevity we write the answer only in the equal mass limit. The divergent piece of the self energy correction, in the equal mass limit, is given by  
\begin{equation}\label{self_energy_corr_equal_mass}
    \begin{split}
        &-(m-\slashed{k}/2)\Big[(y+y_\rho^\ast)(y+y_\sigma+2i\text{Im}[y_\kappa])+(y+y_\kappa^\ast)(y+y_\sigma+2i\text{Im}[y_\rho])\Big]\frac{1}{d-4}\frac{i}{(4\pi)^2}\\
        &-\frac{\slashed{k}}{k^2}\Big[(y_\sigma-y_\rho)(y+y_\sigma+2i\text{Im}[y_\kappa])+(y_\sigma-y_\kappa)(y+y_\sigma+2i\text{Im}[y_\rho])\Big]\frac{1}{d-4}\frac{i(m^2)}{(4\pi)^2} + ...\,.
    \end{split}
\end{equation}
The first line of the above result is local in external momentum. This term comes from the scalar $B$-type contribution that is present inside PV reduction (given in eqn \eqref{zero_gamma5_FS no cut}, eqn \eqref{zero_gamma5_FS two cut} and in eqn \eqref{zero_gamma5_FS one cut}). The second line of eqn \eqref{self_energy_corr_equal_mass} is the non-local divergent piece in the self energy correction. The $A$- type integrals that are present in the PV reduction (given in eqn \eqref{zero_gamma5_FS no cut}, in eqn \eqref{zero_gamma5_FS two cut} and in eqn \eqref{zero_gamma5_FS one cut}), contribute to this non-local divergence. 

If we impose the tree level Lindblad conditions given in eqn \eqref{Lindblad_conditions} in the above one-loop correction, even then the non-local divergences do not cancel in eqn \eqref{self_energy_corr_equal_mass}. So, we find that the non-local divergences persist in one-loop correction to the fermion self-energy in open-Yukawa theory.

It is shown in eqn \cite{Avinash:2017asn} that non-local divergences disappear in the self energy correction to the scalar if the masses are chosen to be equal. So, we find that the divergence structure of the self energy correction to the fermion in the equal mass limit is a distinguishing feature of open Yukawa theory from that of the scalar field theory.

\section{Conclusion}

\label{sec:bjrconclusion}

It was shown in  \cite{Avinash:2017asn} that the $B$-type integrals do not have non-local divergences in open-$\phi^3+\phi^4$ theory. In this paper, we have considered PV tensor reduction of one loop integrals for local quantum field theories on the SK contour. The key philosophy of PV reduction is that, in a general open-QFT, the loop diagrams can be written as a linear combination of scalar loop integrals; the coefficients are non-local functions of external momenta. These non-local contributions seem to prevail in the divergent scalar loop integrals, unlike in unitary QFTs. So, we expect non-local divergences in open-QFTs in general, just by examining the PV reduction formula. However, there is a possibility that the non-local divergences cancel when we sum over all the Feynman diagrams contributing a process. So to infer that a physical observable retains the non-local divergences, we need to do a more careful analysis. For this reason, we consider open Yukawa theory (and also two scalars theory in appendix \ref{app:twoscalar}). We have shown that the open-Yukawa theory possesses non-local divergence in the correction to the self-energy correction to fermion.  

The physical origin and removal of these non-local divergences remain an open question.

\paragraph{Acknowledgement} 

We are thankful to R Loganayagam for collaborating at the early stage of the project.  A.R. would like to thank ICTS-TIFR, Bengaluru for hospitality during this work. C.J. would like to thank ICTP, Trieste for hospitality during this project. We are grateful to   Diksha Jain, R Loganayagam, Akhil Sivakumar especially Joydeep Chakravarty for various suggestions to improve the draft.

\appendix

\section{Feynman diagrams of the paper}
\label{sec:feynmandiagram}

In this appendix, we have depicted Feynman diagrams for various one loop corrections that we have computed in this paper.

\subsection{Scalar tadpole diagrams}
\label{subsec:scalartadpolediagram}

In section \ref{subsec:bjropenyukawascalartadpole}, we have computed one-loop contribution to the scalar tadpole. The corresponding Feynman diagrams can be found here. 

\begin{figure}[ht]
\begin{center}
\begin{tikzpicture}[line width=1 pt, scale=.6]

\begin{scope}[shift={(0,0)}, rotate=90]
\draw [phir, ultra thick, domain=0:360] plot ({1*cos(\x)}, {1*sin(\x)});
\draw [phir, ultra thick] (1,0) -- (2.5,0);
\node at (1,0) {$\times $};
\node at (-2,0) {$ \frac{(-i\lambda_3)}{2} A_{R}$};
\end{scope}

\begin{scope}[shift={(5,0)}, rotate=90]
\draw [phil, ultra thick, domain=0:360] plot ({1*cos(\x)}, {1*sin(\x)});
\draw [phir, ultra thick] (1,0) -- (2.5,0);
\node at (1,0) {$\times $};
\node at (-2,0) {$\frac{(i\lambda_{3\sigma}^\star)}{2}A_{L}$};
\end{scope}

\begin{scope}[shift={(10,0)}, rotate=90] 
\draw [phir, ultra thick, domain=0:180] plot ({1*cos(\x)}, {1*sin(\x)});
\draw [phil, ultra thick, domain=180:360] plot ({1*cos(\x)}, {1*sin(\x)});
\draw [phir, ultra thick] (1,0) -- (2.5,0);
\node at (1,0) {$\times $};
\node at (-2,0) {$(-i\lambda_{3\sigma}) A_{P}$};
\end{scope}

%
\begin{scope}[shift={(-2,-5)}]

\begin{scope}[shift={(0,0)}, rotate=90]
\drawpsidiagar{0}{0}{0}
\draw [phir, ultra thick] (1,0) -- (2.5,0);
\node at (1,0) {$\times $};
\node at (-2,0) {$(-iy)A^f_{R}$};
\end{scope}

\begin{scope}[shift={(5,0)}, rotate=90]
\drawpsidiagal{0}{0}{0}
\draw [phir, ultra thick] (1,0) -- (2.5,0);
\node at (1,0) {$\times $};
\node at (-2,0) {$ (iy_{\sigma}^\star)A^f_{L}$};
\end{scope}

\begin{scope}[shift={(10,0)}, rotate=90] 
\drawpsidiagam{0}{0}{0}
\draw [phir, ultra thick] (1,0) -- (2.5,0);
\node at (1,0) {$\times $};
\node at (-2,0) {$(-iy_\kappa)A^f_{P}$};
\end{scope} 

\begin{scope}[shift={(15,0)}, rotate=90] 
\drawpsidiagap{0}{0}{0}
\draw [phir, ultra thick] (1,0) -- (2.5,0);
\node at (1,0) {$\times $};
\node at (-2,0) {$(-iy_\rho)A^f_{M}$};
\end{scope}

\end{scope}

\end{tikzpicture}
\end{center}
\caption{$\phir$ one loop tadpole }
\label{fig:philtadpole1}
\end{figure}
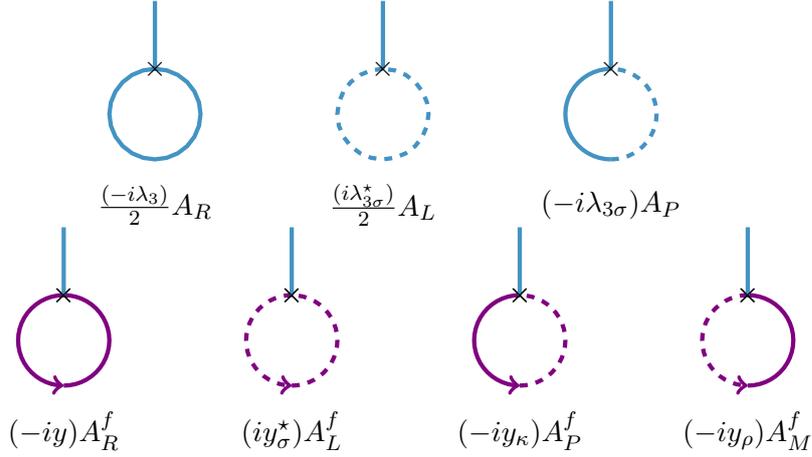

\subsection{Feynman diagrams for scalar mass renormalization}
\label{subsec:scalarmassdiagram}
\begin{figure}[h]
\begin{center}
\begin{tikzpicture}[line width=1 pt, scale=0.6]
 
\begin{scope}[shift={(0,0)}]
\draw [phir, ultra thick] (1,0) -- (2,0);
\draw [phir, ultra thick] (-1,0) -- (-2,0);
\draw [phir, ultra thick, domain=0:180] plot ({1*cos(\x)}, {1*sin(\x)});
\draw [phir, ultra thick, domain=180:360] plot ({1*cos(\x)}, {1*sin(\x)}); 
\node at (-1,0) {$\times $};    
\node at (1,0) {$\times $};
\node at (0,-2.0) {$\frac{(-i\lambda_3)^2}{2} B_{RR}(k)$};    
\node at (-1,0) {$\times $};    


\end{scope}

\begin{scope}[shift={(5,0)}]
\draw [phir, ultra thick] (1,0) -- (2,0);
\draw [phir, ultra thick] (-1,0) -- (-2,0);
\draw [phil, ultra thick, domain=0:90] plot ({1*cos(\x)}, {1*sin(\x)});
\draw [phil, ultra thick, domain=90:180] plot ({1*cos(\x)}, {1*sin(\x)});
\draw [phil, ultra thick, domain=180:270] plot ({1*cos(\x)}, {1*sin(\x)});
\draw [phil, ultra thick, domain=270:360] plot ({1*cos(\x)}, {1*sin(\x)}); 
\node at (-1,0) {$\times $};    
\node at (1,0) {$\times $};    
\node at (0,-2) {$\frac{(i\sigma_3^\star)^2}{2} B_{LL}(k)$};

    
\end{scope}

\begin{scope}[shift={(10,0)}]
\draw [phir, ultra thick] (1,0) -- (2,0);
\draw [phir, ultra thick] (-1,0) -- (-2,0);
\draw [phir, ultra thick, domain=0:90] plot ({1*cos(\x)}, {1*sin(\x)});
\draw [phir, ultra thick, domain=90:180] plot ({1*cos(\x)}, {1*sin(\x)});
\draw [phil, ultra thick, domain=180:270] plot ({1*cos(\x)}, {1*sin(\x)});
\draw [phil, ultra thick, domain=270:360] plot ({1*cos(\x)}, {1*sin(\x)}); 
\node at (-1,0) {$\times $};    
\node at (1,0) {$\times $};    
\node at (0,-2) {$(-i\sigma_3)^2 B_{LR}(k)$};    

    
\end{scope}

\begin{scope}[shift={(15,0)}]
\draw [phir, ultra thick] (1,0) -- (2,0);
\draw [phir, ultra thick] (-1,0) -- (-2,0);
\draw [phil, ultra thick, domain=0:90] plot ({1*cos(\x)}, {1*sin(\x)});
\draw [phir, ultra thick, domain=90:180] plot ({1*cos(\x)}, {1*sin(\x)});
\draw [phir, ultra thick, domain=180:270] plot ({1*cos(\x)}, {1*sin(\x)});
\draw [phil, ultra thick, domain=270:360] plot ({1*cos(\x)}, {1*sin(\x)});  
\node at (-1,0) {$\times $};    
\node at (1,0) {$\times $};    
\node at (0,-2) {$\frac{(i\sigma_3^\star)(-i\lambda_3)}{2} B_{PM}(k)$};    

    
\end{scope}

\begin{scope}[shift={(0,-4)}]
\draw [phir, ultra thick] (1,0) -- (2,0);
\draw [phir, ultra thick] (-1,0) -- (-2,0);
\draw [phir, ultra thick, domain=0:90] plot ({1*cos(\x)}, {1*sin(\x)});
\draw [phil, ultra thick, domain=90:180] plot ({1*cos(\x)}, {1*sin(\x)});
\draw [phil, ultra thick, domain=180:270] plot ({1*cos(\x)}, {1*sin(\x)});
\draw [phir, ultra thick, domain=270:360] plot ({1*cos(\x)}, {1*sin(\x)}); 
\node at (-1,0) {$\times $};    
\node at (1,0) {$\times $};    
\node at (0,-2) {$\frac{(i\sigma_3^\star)(-i\lambda_3)}{2} B_{MP}(k)$};    

    
\end{scope}

\begin{scope}[shift={(5,-4)}]
\draw [phir, ultra thick] (1,0) -- (2,0);
\draw [phir, ultra thick] (-1,0) -- (-2,0);
\draw [phir, ultra thick, domain=0:90] plot ({1*cos(\x)}, {1*sin(\x)});
\draw [phil, ultra thick, domain=90:180] plot ({1*cos(\x)}, {1*sin(\x)});
\draw [phir, ultra thick, domain=180:270] plot ({1*cos(\x)}, {1*sin(\x)});
\draw [phil, ultra thick, domain=270:360] plot ({1*cos(\x)}, {1*sin(\x)}); 
\node at (-1,0) {$\times $};    
\node at (1,0) {$\times $};    
\node at (0,-2) {$(-i\sigma_3)^2 B_{PP}(k)$};    
    

\end{scope}

\begin{scope}[shift={(10,-4)}]
\draw [phir, ultra thick] (1,0) -- (2,0);
\draw [phir, ultra thick] (-1,0) -- (-2,0);
\draw [phir, ultra thick, domain=0:90] plot ({1*cos(\x)}, {1*sin(\x)});
\draw [phir, ultra thick, domain=90:180] plot ({1*cos(\x)}, {1*sin(\x)});
\draw [phir, ultra thick, domain=180:270] plot ({1*cos(\x)}, {1*sin(\x)});
\draw [phil, ultra thick, domain=270:360] plot ({1*cos(\x)}, {1*sin(\x)}); 
\node at (-1,0) {$\times $};     
\node at (1,0) {$\times $};    
\node at (0,-2) {$-(\lambda_3\sigma_3) B_{PR}(k)$};    

    
\end{scope}

\begin{scope}[shift={(15,-4)}]
\draw [phir, ultra thick] (1,0) -- (2,0);
\draw [phir, ultra thick] (-1,0) -- (-2,0);
\draw [phir, ultra thick, domain=0:90] plot ({1*cos(\x)}, {1*sin(\x)});
\draw [phir, ultra thick, domain=90:180] plot ({1*cos(\x)}, {1*sin(\x)});
\draw [phil, ultra thick, domain=180:270] plot ({1*cos(\x)}, {1*sin(\x)});
\draw [phir, ultra thick, domain=270:360] plot ({1*cos(\x)}, {1*sin(\x)}); 
\node at (-1,0) {$\times $};    
\node at (1,0) {$\times $};    
\node at (0,-2) {$-(\lambda_3\sigma_3) B_{MR}(k)$};

    
\end{scope}

\begin{scope}[shift={(4,-8)}]
\draw [phir, ultra thick] (1,0) -- (2,0);
\draw [phir, ultra thick] (-1,0) -- (-2,0);
\draw [phil, ultra thick, domain=0:90] plot ({1*cos(\x)}, {1*sin(\x)});
\draw [phil, ultra thick, domain=90:180] plot ({1*cos(\x)}, {1*sin(\x)});
\draw [phir, ultra thick, domain=180:270] plot ({1*cos(\x)}, {1*sin(\x)});
\draw [phil, ultra thick, domain=270:360] plot ({1*cos(\x)}, {1*sin(\x)}); 
\node at (-1,0) {$\times $};    
\node at (1,0) {$\times $};    
\node at (0,-2) {$(i\sigma_3^\star)(-i\sigma_3) B_{PL} (k)$};    

    
\end{scope}

\begin{scope}[shift={(11,-8)}]
\draw [phir, ultra thick] (1,0) -- (2,0);
\draw [phir, ultra thick] (-1,0) -- (-2,0);
\draw [phil, ultra thick, domain=0:90] plot ({1*cos(\x)}, {1*sin(\x)});
\draw [phil, ultra thick, domain=90:180] plot ({1*cos(\x)}, {1*sin(\x)});
\draw [phil, ultra thick, domain=180:270] plot ({1*cos(\x)}, {1*sin(\x)});
\draw [phir, ultra thick, domain=270:360] plot ({1*cos(\x)}, {1*sin(\x)});  
\node at (-1,0) {$\times $};    
\node at (1,0) {$\times $};    
\node at (0,-2) {$(i\sigma_3^\star)(-i\sigma_3) B_{ML}(k)$};    

    
\end{scope}

\end{tikzpicture}
\end{center}
\caption{One loop corrections to $ m^2_\phi$ due to cubic couplings}
\label{diag:macromassrenorm1}
\end{figure}
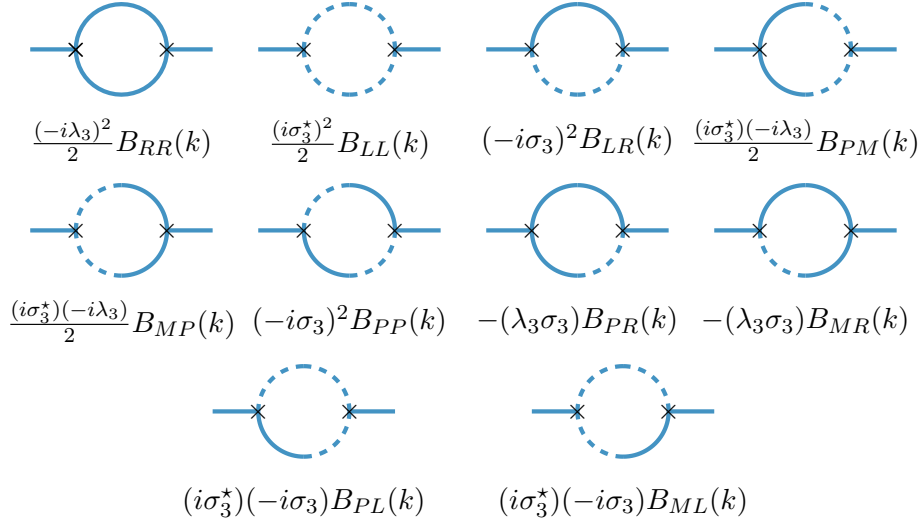

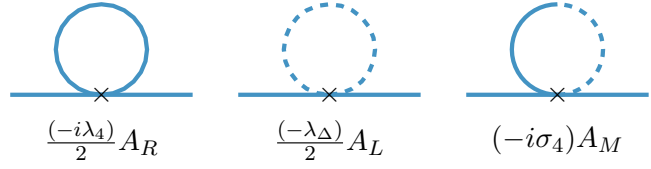
\begin{figure}[ht]
\begin{center}
\begin{tikzpicture}[line width=1 pt, scale=0.6]

\begin{scope}[shift={(0,0)}, rotate=-90]
\draw [phir, ultra thick, domain=0:360] plot ({1*cos(\x)}, {1*sin(\x)});
\draw [phir, ultra thick] (1,0) -- (1,2);
\draw [phir, ultra thick] (1,0) -- (1,-2);
\node at (1,0) {$\times $};
\node at (2,0) {$ \frac{(-i\lambda_4)}{2} A_{R}$};
\end{scope}

\begin{scope}[shift={(5,0)}, rotate=-90]
\draw [phil, ultra thick, domain=0:360] plot ({1*cos(\x)}, {1*sin(\x)});
\draw [phir, ultra thick] (1,0) -- (1,2);
\draw [phir, ultra thick] (1,0) -- (1,-2);
\node at (1,0) {$\times $};
\node at (2,0) {$\frac{(-\lambda_\Delta)}{2}A_{L}$};
\end{scope}

\begin{scope}[shift={(10,0)}, rotate=-90] 
\draw [phil, ultra thick, domain=0:180] plot ({1*cos(\x)}, {1*sin(\x)});
\draw [phir, ultra thick, domain=180:360] plot ({1*cos(\x)}, {1*sin(\x)});
\draw [phir, ultra thick] (1,0) -- (1,2);
\draw [phir, ultra thick] (1,0) -- (1,-2);
\node at (1,0) {$\times $};
\node at (2,0) {$(-i\sigma_4) A_{M}$};
\end{scope}

\end{tikzpicture}
\end{center}
\caption{One loop correction to $m^2_\phi$ due to quartic couplings}
\label{fig:macromassrenorm2}
\end{figure}

The one loop correction to scalar mass was computed in section \ref{subsec:bjropenyukawascalarmass}. Three different kinds of one loop diagrams contribute to the self energy correction of the scalar field - scalar tadpole, scalar bubble and fermionic bubble. The corresponding the Feynman can be found here. The scalar bubble diagrams are depicted 
 in fig. \ref{diag:macromassrenorm1},  the scalar tadpole diagrams are depicted in fig. \ref{fig:macromassrenorm2} and the fermionic bubble  diagrams are shown in  fig. \ref{diag:macromassrenorm3} and in fig. \ref{diag:macromassrenorm4}.

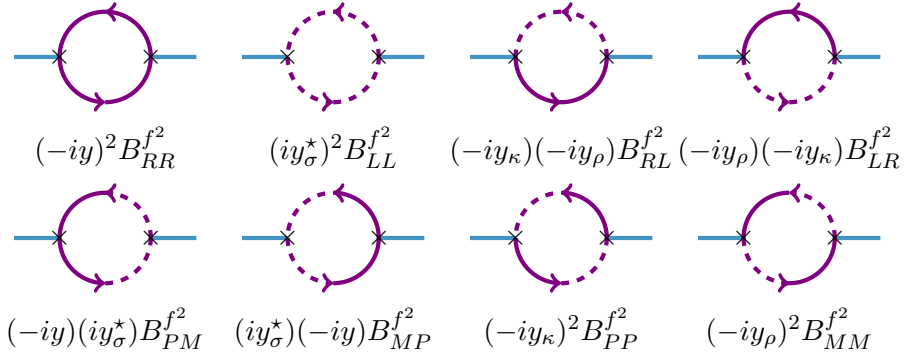
\begin{figure}[h]
\begin{center}
\begin{tikzpicture}[line width=1 pt, scale=0.6]
 
\begin{scope}[shift={(0,0)}]
\draw [phir, ultra thick] (1,0) -- (2,0);
\draw [phir, ultra thick] (-1,0) -- (-2,0);
\drawpsidiagbrr{0}{0}{0}
\node at (0,-2) { $(-iy)^2 {B}^{f^2}_{RR}$};


\end{scope}

\begin{scope}[shift={(5,0)}]
\draw [phir, ultra thick] (1,0) -- (2,0);
\draw [phir, ultra thick] (-1,0) -- (-2,0);
\drawpsidiagbll{0}{0}{0}     
\node at (0,-2) {$(iy_\sigma^\star)^2 {B}^{f^2}_{LL}$};
    
\end{scope}
 
\begin{scope}[shift={(10,0)}]
\draw [phir, ultra thick] (1,0) -- (2,0);
\draw [phir, ultra thick] (-1,0) -- (-2,0);
\drawpsidiagblr{0}{0}{180}    
\node at (0,-2) {$(-iy_\kappa)(-iy_\rho) {B}^{f^2}_{RL}$};    

    
\end{scope}

\begin{scope}[shift={(15,0)}]
\draw [phir, ultra thick] (1,0) -- (2,0);
\draw [phir, ultra thick] (-1,0) -- (-2,0);
\drawpsidiagblr{0}{0}{0}    
\node at (0,-2) {$(-iy_\rho)(-iy_\kappa) {B}^{f^2}_{LR}$};    

    
\end{scope}

\begin{scope}[shift={(0,-4)}]
\draw [phir, ultra thick] (1,0) -- (2,0);
\draw [phir, ultra thick] (-1,0) -- (-2,0);
\drawpsidiagbpm{0}{0}{0}    
\node at (0,-2) {$(-iy)(iy_\sigma^\star) {B}^{f^2}_{PM}$};    

    
\end{scope}

\begin{scope}[shift={(5,-4)}]
\draw [phir, ultra thick] (1,0) -- (2,0);
\draw [phir, ultra thick] (-1,0) -- (-2,0);
\drawpsidiagbmp{0}{0}{0}
\node at (0,-2) {$(iy_\sigma^\star)(-iy) {B}^{f^2}_{MP}$};    

     
\end{scope}

\begin{scope}[shift={(10,-4)}]
\draw [phir, ultra thick] (1,0) -- (2,0);
\draw [phir, ultra thick] (-1,0) -- (-2,0);
\drawpsidiagbpp{0}{0}{0}
\node at (0,-2) {$(-iy_\kappa)^2 {B}^{f^2}_{PP}$};    
    

\end{scope}

\begin{scope}[shift={(15,-4)}]
\draw [phir, ultra thick] (1,0) -- (2,0);
\draw [phir, ultra thick] (-1,0) -- (-2,0);
\drawpsidiagbmm{0}{0}{180}
\node at (0,-2) {$(-iy_\rho)^2 {B}^{f^2}_{MM}$};    
    

\end{scope}
\end{tikzpicture}
\end{center}
 \caption{One loop corrections to $ m^2_\phi$ due to fermionic couplings}
 \label{diag:macromassrenorm3}
\end{figure}

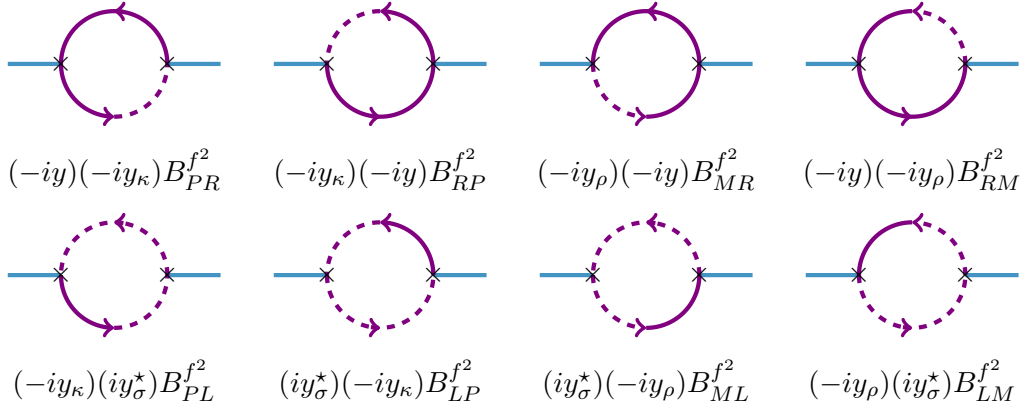
\begin{figure}
\begin{center}
\begin{tikzpicture}[line width=1 pt, scale=0.7]

\begin{scope}[shift={(0,0)}]
\draw [phir, ultra thick] (1,0) -- (2,0);
\draw [phir, ultra thick] (-1,0) -- (-2,0);
\drawpsidiagbpr{0}{0}{0} 

\node at (0,-2) {$(-iy)(-iy_\kappa) {B}^{f^2}_{PR}$};    

    
\end{scope}

\begin{scope}[shift={(5,0)}]
\draw [phir, ultra thick] (1,0) -- (2,0);
\draw [phir, ultra thick] (-1,0) -- (-2,0);
\drawpsidiagbpr{0}{0}{180} 

\node at (0,-2) {$(-iy_\kappa)(-iy) {B}^{f^2}_{RP}$};    

    
\end{scope}

\begin{scope}[shift={(10,0)}]
\draw [phir, ultra thick] (1,0) -- (2,0);
\draw [phir, ultra thick] (-1,0) -- (-2,0);
\drawpsidiagbmr{0}{0}{0}

\node at (0,-2) {$(-iy_\rho)(-iy) {B}^{f^2}_{MR}$};

    
\end{scope}

\begin{scope}[shift={(15,0)}]
\draw [phir, ultra thick] (1,0) -- (2,0);
\draw [phir, ultra thick] (-1,0) -- (-2,0);
\drawpsidiagbmr{0}{0}{180}

\node at (0,-2) {$(-iy)(-iy_\rho) {B}^{f^2}_{RM}$};

    
\end{scope}

\begin{scope}[shift={(0,-4)}]
\draw [phir, ultra thick] (1,0) -- (2,0);
\draw [phir, ultra thick] (-1,0) -- (-2,0);
\drawpsidiagbpl{0}{0}{0}

\node at (0,-2) {$(-iy_\kappa)(iy_\sigma^\star) {B}^{f^2}_{PL}$};    

    
\end{scope}

\begin{scope}[shift={(5,-4)}]
\draw [phir, ultra thick] (1,0) -- (2,0);
\draw [phir, ultra thick] (-1,0) -- (-2,0);
\drawpsidiagbpl{0}{0}{180}

\node at (0,-2) {$(iy_\sigma^\star)(-iy_\kappa) {B}^{f^2}_{LP}$};    

    
\end{scope}

\begin{scope}[shift={(10,-4)}]
\draw [phir, ultra thick] (1,0) -- (2,0);
\draw [phir, ultra thick] (-1,0) -- (-2,0);
\drawpsidiagbml{0}{0}{0}

\node at (0,-2) {$(iy_\sigma^\star)(-iy_\rho) {B}^{f^2}_{ML}$};    


\end{scope}     

\begin{scope}[shift={(15,-4)}]
\draw [phir, ultra thick] (1,0) -- (2,0);
\draw [phir, ultra thick] (-1,0) -- (-2,0);
\drawpsidiagbml{0}{0}{180}

\node at (0,-2) {$(-iy_\rho)(iy_\sigma^\star) {B}^{f^2}_{LM}$};    


\end{scope}

\end{tikzpicture}
\end{center}
\caption{One loop corrections to $ m^2_\phi$ due to fermionic couplings}
\label{diag:macromassrenorm4}
\end{figure}

\newpage
\subsection{Feynman diagrams for fermionic mass renormalization}
\label{subsec:fermionicmassdiagram}

\begin{figure}[h]
\begin{center}
\begin{tikzpicture}[scale=0.6]
\begin{scope}[shift={(0,0)}]
\draw[ultra thick, violet,->] (-1,0) arc (180:270:1);
\draw[ultra thick, violet] (0,-1) arc (270:360:1);
\draw[ultra thick, blue] (1,0) arc (0:90:1);
\draw[ultra thick, blue] (0,1) arc (90:180:1);
\draw[ultra thick, violet] (-2,0) -- (-1,0);
\draw[ultra thick, violet] (2,0) -- (1,0);
\node at (0,-1.6) {$ (-iy)^2 B^{f}_{RR}$};
\end{scope}

\begin{scope}[shift={(5,0)}]
\draw[ultra thick, violet,dashed,->] (-1,0) arc (180:270:1);
\draw[ultra thick, violet,dashed] (0,-1) arc (270:360:1);
\draw[ultra thick, blue, dashed] (1,0) arc (0:90:1);
\draw[ultra thick, blue, dashed] (0,1) arc (90:180:1);
\draw[ultra thick, violet] (-2,0) -- (-1,0);
\draw[ultra thick, violet] (2,0) -- (1,0);
\node at (0,-1.6) {$(iy_\rho^\ast)(iy_\kappa^\ast) B^{f}_{LL}$};
\end{scope}

\begin{scope}[shift={(10,0)}]
\draw[ultra thick, violet,->] (-1,0) arc (180:270:1);
\draw[ultra thick, violet] (0,-1) arc (270:360:1);
\draw[ultra thick, blue, dashed] (1,0) arc (0:90:1);
\draw[ultra thick, blue, dashed] (0,1) arc (90:180:1);
\draw[ultra thick, violet] (-2,0) -- (-1,0);
\draw[ultra thick, violet] (2,0) -- (1,0);
\node at (0,-1.6) {$ (-iy_\sigma)^2 B^{f}_{RL}$};
\end{scope}

\begin{scope}[shift={(15,0)}]
\draw[ultra thick, violet, dashed,->] (-1,0) arc (180:270:1);
\draw[ultra thick, violet, dashed] (0,-1) arc (270:360:1);
\draw[ultra thick, blue] (1,0) arc (0:90:1);
\draw[ultra thick, blue] (0,1) arc (90:180:1);
\draw[ultra thick, violet] (-2,0) -- (-1,0);
\draw[ultra thick, violet] (2,0) -- (1,0);
\node at (0,-1.6) {$ (-iy_\rho)(-iy_\kappa) B^{f}_{LR}$};
\end{scope}

\begin{scope}[shift={(0,-4)}]
\draw[ultra thick, violet,->] (-1,0) arc (180:270:1);
\draw[ultra thick, violet] (0,-1) arc (270:360:1);
\draw[ultra thick, blue] (1,0) arc (0:90:1);
\draw[ultra thick, blue, dashed] (0,1) arc (90:180:1);
\draw[ultra thick, violet] (-2,0) -- (-1,0);
\draw[ultra thick, violet] (2,0) -- (1,0);
\node at (0,-1.6) {$(-iy_\sigma)(-iy) B^{f}_{RP}$};
\end{scope}

\begin{scope}[shift={(5,-4)}]
\draw[ultra thick, violet,->] (-1,0) arc (180:270:1);
\draw[ultra thick, violet] (0,-1) arc (270:360:1);
\draw[ultra thick, blue, dashed] (1,0) arc (0:90:1);
\draw[ultra thick, blue] (0,1) arc (90:180:1);
\draw[ultra thick, violet] (-2,0) -- (-1,0);
\draw[ultra thick, violet] (2,0) -- (1,0);
\node at (0,-1.6) {$ (-iy)(-iy_\sigma) B^{f}_{RM}$};
\end{scope}

\begin{scope}[shift={(10,-4)}]
\draw[ultra thick, violet,->] (-1,0) arc (180:270:1);
\draw[ultra thick, violet, dashed] (0,-1) arc (270:360:1);
\draw[ultra thick, blue] (1,0) arc (0:90:1);
\draw[ultra thick, blue] (0,1) arc (90:180:1);
\draw[ultra thick, violet] (-2,0) -- (-1,0);
\draw[ultra thick, violet] (2,0) -- (1,0);
\node at (0,-1.6) {$ (-iy)(-iy_\kappa) B^{f}_{PR}$};
\end{scope}

\begin{scope}[shift={(15,-4)}]
\draw[ultra thick, violet, dashed,->] (-1,0) arc (180:270:1);
\draw[ultra thick, violet] (0,-1) arc (270:360:1);
\draw[ultra thick, blue] (1,0) arc (0:90:1);
\draw[ultra thick, blue] (0,1) arc (90:180:1);
\draw[ultra thick, violet] (-2,0) -- (-1,0);
\draw[ultra thick, violet] (2,0) -- (1,0);
\node at (0,-1.6) {$ (-iy_\rho)(-iy) B^{f}_{MR}$};
\end{scope}


\begin{scope}[shift={(0,-8)}]
\draw[ultra thick, violet, dashed,->] (-1,0) arc (180:270:1);
\draw[ultra thick, violet, dashed] (0,-1) arc (270:360:1);
\draw[ultra thick, blue] (1,0) arc (0:90:1);
\draw[ultra thick, blue, dashed] (0,1) arc (90:180:1);
\draw[ultra thick, violet] (-2,0) -- (-1,0);
\draw[ultra thick, violet] (2,0) -- (1,0);
\node at (0,-1.6) {$ (iy_\rho^\ast)(-iy_\kappa) B^{f}_{LP}$};
\end{scope}

\begin{scope}[shift={(5,-8)}]
\draw[ultra thick, violet, dashed,->] (-1,0) arc (180:270:1);
\draw[ultra thick, violet, dashed] (0,-1) arc (270:360:1);
\draw[ultra thick, blue, dashed] (1,0) arc (0:90:1);
\draw[ultra thick, blue] (0,1) arc (90:180:1);
\draw[ultra thick, violet] (-2,0) -- (-1,0);
\draw[ultra thick, violet] (2,0) -- (1,0);
\node at (0,-1.6) {$ (-iy_\rho)(iy_\kappa^\ast) B^{f}_{LM}$};
\end{scope}

\begin{scope}[shift={(10,-8)}]
\draw[ultra thick, violet,->] (-1,0) arc (180:270:1);
\draw[ultra thick, violet, dashed] (0,-1) arc (270:360:1);
\draw[ultra thick, blue, dashed] (1,0) arc (0:90:1);
\draw[ultra thick, blue, dashed] (0,1) arc (90:180:1);
\draw[ultra thick, violet] (-2,0) -- (-1,0);
\draw[ultra thick, violet] (2,0) -- (1,0);
\node at (0,-1.6) {$ (-iy_\sigma)(iy_\kappa^\ast) B^{f}_{PL}$};
\end{scope}

\begin{scope}[shift={(15,-8)}]
\draw[ultra thick, violet, dashed,->] (-1,0) arc (180:270:1);
\draw[ultra thick, violet] (0,-1) arc (270:360:1);
\draw[ultra thick, blue, dashed] (1,0) arc (0:90:1);
\draw[ultra thick, blue, dashed] (0,1) arc (90:180:1);
\draw[ultra thick, violet] (-2,0) -- (-1,0);
\draw[ultra thick, violet] (2,0) -- (1,0);
\node at (0,-1.6) {$ (iy_\rho^\ast)(-iy_\sigma) B^{f}_{ML}$};
\end{scope}


\begin{scope}[shift={(0,-12)}]
\draw[ultra thick, violet,->] (-1,0) arc (180:270:1);
\draw[ultra thick, violet, dashed] (0,-1) arc (270:360:1);
\draw[ultra thick, blue] (1,0) arc (0:90:1);
\draw[ultra thick, blue, dashed] (0,1) arc (90:180:1);
\draw[ultra thick, violet] (-2,0) -- (-1,0);
\draw[ultra thick, violet] (2,0) -- (1,0);
\node at (0,-1.6) {$ (-iy_\rho)(-iy_\kappa) B^{f}_{PP}$};
\end{scope}

\begin{scope}[shift={(5,-12)}]
\draw[ultra thick, violet,->] (-1,0) arc (180:270:1);
\draw[ultra thick, violet, dashed] (0,-1) arc (270:360:1);
\draw[ultra thick, blue, dashed] (1,0) arc (0:90:1);
\draw[ultra thick, blue] (0,1) arc (90:180:1);
\draw[ultra thick, violet] (-2,0) -- (-1,0);
\draw[ultra thick, violet] (2,0) -- (1,0);
\node at (0,-1.6) {$ (-iy)(iy_\kappa^\ast) B^{f}_{PM}$};
\end{scope}

\begin{scope}[shift={(10,-12)}]
\draw[ultra thick, violet, dashed,->] (-1,0) arc (180:270:1);
\draw[ultra thick, violet] (0,-1) arc (270:360:1);
\draw[ultra thick, blue] (1,0) arc (0:90:1);
\draw[ultra thick, blue, dashed] (0,1) arc (90:180:1);
\draw[ultra thick, violet] (-2,0) -- (-1,0);
\draw[ultra thick, violet] (2,0) -- (1,0);
\node at (0,-1.6) {$ (iy_\rho^\ast)(-iy) B^{f}_{MP}$};
\end{scope}

\begin{scope}[shift={(15,-12)}]
\draw[ultra thick, violet, dashed,->] (-1,0) arc (180:270:1);
\draw[ultra thick, violet] (0,-1) arc (270:360:1);
\draw[ultra thick, blue, dashed] (1,0) arc (0:90:1);
\draw[ultra thick, blue] (0,1) arc (90:180:1);
\draw[ultra thick, violet] (-2,0) -- (-1,0);
\draw[ultra thick, violet] (2,0) -- (1,0);
\node at (0,-1.6) {$ (-iy_\rho)(-iy_\sigma) B^{f}_{MM}$};
\end{scope}

\end{tikzpicture}
\end{center}
\caption{One loop mass renormalization of the fermion}
\label{fig:fermionselfenergy}
\end{figure}
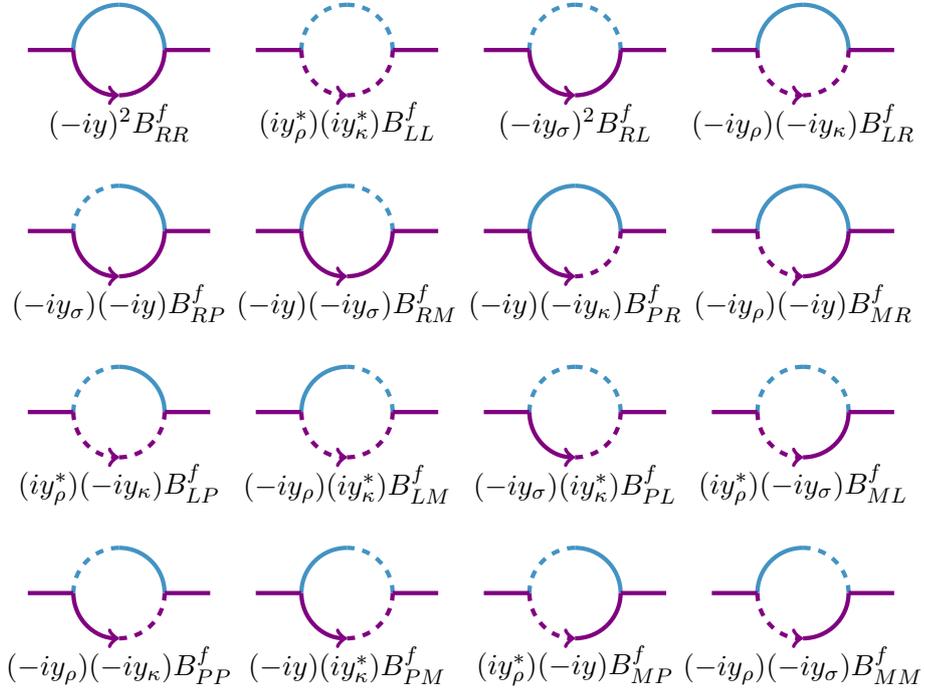

We computed the one loop correction to fermionic mass in section \ref{subsec:bjropenyukawafermionmass}. The Feynman diagrams which contribute to fermionic mass renormalization are depicted in  fig. \ref{fig:fermionselfenergy}. 

\newpage

\section{PV reduction of $B^{f}_{ab}(k,m_{\psi},m_{\phi})$ in open Yukawa theory}
\label{sec:fermionicpv}

In this section we write the PV reduction for all $B^f_{ab}$ loop integrals. The $f$ superscript denotes that one propagator is fermionic and the other is scalar. In order to write the PV reduction formulae, it is be convenient to define the following quantity 
\begin{eqnarray}
\Delta \equiv-k^2+m_\psi^2-m_\phi^2\,,
\end{eqnarray}
The PV reduction formula for $B^f_{ab}$ loop integrals are given below. The reduction formula for $B^f$ integrals with no cut propagator are given by 
\begin{equation}
\label{zero_gamma5_FS no cut}
\begin{split}
B^{f}_{RR}(k,m_{\psi},m_{\phi}) &= m_{\psi} B_{RR}\left(k,m_{\psi},m_{\phi}\right) -\slashed{k}\ \frac{\ i \  A_R\left(m_{\psi}\right) -\ i \ A_R \left(m_{\phi}\right)-\Delta B_{RR}\left(k,m_{\psi},m_{\phi}\right)}{2 k^2} \,,\\
B^{f}_{LL}(k,m_{\psi},m_{\phi}) &= m_{\psi} B_{LL}\left(k,m_{\psi},m_{\phi}\right) -\slashed{k}\ \frac{-\ i \ A_L\left(m_{\psi}\right) +\ i \ A_L \left(m_{\phi}\right)-\Delta B_{LL}\left(k,m_{\psi},m_{\phi}\right)}{2 k^2} \,,\\
B^{f}_{RL}(k,m_{\psi},m_{\phi}) &= m_{\psi} B_{RL}\left(k,m_{\psi},m_{\phi}\right) -\slashed{k}\ \frac{-\ i \ A_R\left(m_{\psi}\right) -\ i \ A_L \left(m_{\phi}\right)-\Delta B_{RL}\left(k,m_{\psi},m_{\phi}\right)}{2 k^2} \,,\\
B^{f}_{LR}(k,m_{\psi},m_{\phi}) &= m_{\psi} B_{LR}\left(k,m_{\psi},m_{\phi}\right) -\slashed{k}\ \frac{\ i \ A_R\left(m_{\psi}\right) +\ i \ A_L \left(m_{\phi}\right)-\Delta B_{LR}\left(k,m_{\psi},m_{\phi}\right)}{2 k^2} \,,
\end{split}
\end{equation}
Now  consider reduction formula for $B^f$ with two cut propagator. They are given by 
\begin{equation}\label{zero_gamma5_FS two cut}
\begin{split}
B^{f}_{PM}(k,m_{\psi},m_{\phi}) &= m_{\psi} B_{PM}\left(k,m_{\psi},m_{\phi}\right) +\slashed{k}\ \frac{\Delta B_{PM}\left(k,m_{\psi},m_{\phi}\right)}{2 k^2}\,,\\
B^{f}_{MP}(k,m_{\psi},m_{\phi}) &= m_{\psi} B_{MP}\left(k,m_{\psi},m_{\phi}\right) +\slashed{k}\ \frac{\Delta B_{MP}\left(k,m_{\psi},m_{\phi}\right)}{2 k^2}\,,\\
B^{f}_{PP}(k,m_{\psi},m_{\phi}) &= m_{\psi} B_{PP}\left(k,m_{\psi},m_{\phi}\right) +\slashed{k}\ \frac{\Delta B_{PP}\left(k,m_{\psi},m_{\phi}\right)}{2 k^2}\,,\\
B^{f}_{MM}(k,m_{\psi},m_{\phi}) &= m_{\psi} B_{MM}\left(k,m_{\psi},m_{\phi}\right) +\slashed{k}\ \frac{\Delta B_{MM}\left(k,m_{\psi},m_{\phi}\right)}{2 k^2}\,,
\end{split}
\end{equation}
The remaining $B^f$ integrals have one cut propagator. Their PV reduction formulae are given by 
\begin{equation}\label{zero_gamma5_FS one cut}
\begin{split}
B^{f}_{RP}(k,m_{\psi},m_{\phi}) &=m_{\psi} B_{RP}\left(k,m_{\psi},m_{\phi}\right) -\slashed{k}\ \frac{-\ i \ A_P\left(m_{\phi}\right)-\Delta B_{RP}\left(k,m_{\psi},m_{\phi}\right)}{2 k^2} \,,\\
B^{f}_{PR}(k,m_{\psi},m_{\phi}) &=m_{\psi} B_{PR}\left(k,m_{\psi},m_{\phi}\right) -\slashed{k}\ \frac{\ i \ A_P\left(m_{\psi}\right)-\Delta B_{PR}\left(k,m_{\psi},m_{\phi}\right)}{2 k^2} \,,\\
B^{f}_{RM}(k,m_{\psi},m_{\phi}) &= m_{\psi} B_{RM}\left(k,m_{\psi},m_{\phi}\right) -\slashed{k}\ \frac{-\ i \ A_M\left(m_{\phi}\right)-\Delta B_{RM}\left(k,m_{\psi},m_{\phi}\right)}{2 k^2} \,,\\
B^{f}_{MR}(k,m_{\psi},m_{\phi}) &= m_{\psi} B_{MR}\left(k,m_{\psi},m_{\phi}\right) -\slashed{k}\ \frac{\ i \ A_M\left(m_{\psi}\right)-\Delta B_{MR}\left(k,m_{\psi},m_{\phi}\right)}{2 k^2} \,,\\
B^{f}_{LP}(k,m_{\psi},m_{\phi}) &=m_{\psi} B_{LP}\left(k,m_{\psi},m_{\phi}\right) -\slashed{k}\ \frac{\ i \ A_P\left(m_{\phi}\right) -\Delta B_{LP}\left(k,m_{\psi},m_{\phi}\right)}{2 k^2} \,,\\
B^{f}_{PL}(k,m_{\psi},m_{\phi}) &=m_{\psi} B_{PL}\left(k,m_{\psi},m_{\phi}\right) -\slashed{k}\ \frac{-\ i \ A_P\left(m_{\psi}\right) -\Delta B_{PL}\left(k,m_{\psi},m_{\phi}\right)}{2 k^2} \,,\\
B^{f}_{LM}(k,m_{\psi},m_{\phi}) &= m_{\psi} B_{LM}\left(k,m_{\psi},m_{\phi}\right) -\slashed{k}\ \frac{\ i \ A_M\left(m_{\phi}\right) -\Delta B_{LM}\left(k,m_{\psi},m_{\phi}\right)}{2 k^2} \,,\\
B^{f}_{ML}(k,m_{\psi},m_{\phi}) &= m_{\psi} B_{ML}\left(k,m_{\psi},m_{\phi}\right) -\slashed{k}\ \frac{-\ i \ A_M\left(m_{\psi}\right) -\Delta B_{ML}\left(k,m_{\psi},m_{\phi}\right)}{2 k^2} \,.\\
\end{split}
\end{equation} 
If one substitutes the results given in eqn \eqref{scalaratype4} and in the table \eqref{Tab:divergences} in eqn \eqref{zero_gamma5_FS no cut}, eqn \eqref{zero_gamma5_FS two cut} and eqn \eqref{zero_gamma5_FS one cut}, then only $B^f_{RR},B^f_{LL}$ from \eqref{zero_gamma5_FS no cut} and all of the \eqref{zero_gamma5_FS two cut} have no non-local divergences. 

Now if one takes the equal mass limit of the scalar and the fermion ($m_\phi=m_\psi=m$) then one can show that $B_{RL},\,B_{LR}$ in \eqref{zero_gamma5_FS no cut} have non-local divergences and are given by,
\begin{equation}\label{zero_gamma5_FS no cut equal mass}
\begin{split}
B^{f}_{RL}(k,m,m) &= \left(m -\frac{\slashed{k}}{2}\right)\, B_{RL}\left(k,m,m\right) +\frac{\slashed{ik}}{2k^2}\left( A_R(m) +A_L(m)\right) \,,\\
B^{f}_{LR}(k,m,m) &= \left(m -\frac{\slashed{k}}{2}\right)\, B_{LR}\left(k,m,m\right) -\frac{\slashed{ik}}{2k^2}\left( A_R(m) +A_L(m)\right) \,.
\end{split}
\end{equation}
None of the PV reductions in \eqref{zero_gamma5_FS two cut} in the equal mass limit have non-local divergences. All integrals with one cut propagator (given in \eqref{zero_gamma5_FS one cut}) have non-local divergence in their PV reduction formula even in the equal mass limit. We write the divergence for two of them.
\begin{equation}
\label{zero_gamma5_FS one cut equal mass}
\begin{split}
B^{f}_{RP}(k,m,m) &= \left(m -\frac{\slashed{k}}{2}\right)\, B_{RP}\left(k,m,m\right) +\frac{\slashed{ik}}{2k^2}\,A_P(m) \,,\\
B^{f}_{PR}(k,m,m) &=\left(m -\frac{\slashed{k}}{2}\right)\, B_{PR}\left(k,m,m\right) -\frac{\slashed{ik}}{2k^2}\,A_P(m) \,.
\end{split}
\end{equation} 

\section{Open two scalars theory}
\label{app:twoscalar}

Consider a unitary field theory of two real scalar fields ($\phi,\chi$) which interact via $\phi^2\chi^2$ interecting term in the Lagrangian. The Lagrangian is given by, 
 \begin{equation}\label{twoScalarUnitaryLagrangian}
\begin{split}
-\Biggl[ \frac{1}{2} z_\phi\ (\partial \phi)^2 + \frac{1}{2} m_\phi^2  \phi^2+\frac{\lambda_{\phi}}{4!} \phi^4  \Biggr]
-\Biggl[ 
\frac{1}{2} z_\chi\ (\partial \chi)^2 + \frac{1}{2} m_\chi^2  \chi^2+\frac{\lambda_{\chi}}{4!} \chi^4  \Biggr]
-\frac{h}{2!2!}\phi^2\chi^2 
\end{split}
\end{equation}
We have dropped the cubic terms to keep the calculation simple and get the physics. This action has $\mathbb{Z}_2\times \mathbb{Z}_2$ symmetry.
\begin{eqnarray}
\mathbb{Z}_2\times \mathbb{Z}_2
\quad:\quad
\left[
\begin{matrix}
\phi\rightarrow -\phi \qquad& \chi\rightarrow \chi 
\\
\\
\phi\rightarrow \phi \qquad& \chi\rightarrow -\chi 
\end{matrix}
\right. 
\qquad.
\label{z2z2symmetry}
\end{eqnarray}
Since perturbation theory preserves this symmetry, any cubic term won't be generated in perturbative correction.  A point to note that we have kept the masses of the fields to be different and this mass difference will lead us to non-local divergences in the open version of this theory.

Now we write the SK Lagrangian corresponding to \eqref{twoScalarUnitaryLagrangian}. The first two lines of \eqref{twoScalarUnitaryLagrangian} have their own self-interacting SK Lagrangians explained in \cite{Avinash:2017asn} and those are the following
 \begin{equation}
\begin{split}
\mathcal{L}_\phi &
=-\Bigg[ \frac{1}{2} z_\phi\ (\partial \phir)^2 + \frac{1}{2} m_\phi^2  \phir^2+\frac{\lambda_{\phi}}{4!} \phir^4 + \frac{\sigma_{\phi}}{3!} \phir^3 \phil \Bigg]\\
&\qquad+\Bigg[ \frac{1}{2} z_\phi^\star (\partial \phil)^2 + \frac{1}{2} {m_\phi^2}^\star \phil^2 + \frac{\lambda_\phi^\star}{4!} \phil^4+\frac{\sigma_\phi^\star}{3!} \phil^3 \phir\Bigg] \\
&\qquad + i  \Bigg[ z_{\phi\Delta}\ (\partial \phir).(\partial \phil)  + m^2_{\phi\Delta} \phir \phil+\frac{{\lambda_{\phi\Delta}}}{2!2!} \phir^2 \phil^2 \Bigg] 
\end{split}
\end{equation}
and
\begin{equation}
\begin{split}
\mathcal{L}_\chi &
=-\Bigg[ \frac{1}{2} z_\chi\ (\partial \chi_R)^2 + \frac{1}{2} m_\chi^2  \chi_R^2+\frac{\lambda_{\chi}}{4!} \chi_R^4 + \frac{\sigma_{\chi}}{3!} \chi_R^3 \chi_L \Bigg]\\
&\qquad+\Bigg[ \frac{1}{2} z_\chi^\star (\partial \chi_L)^2 + \frac{1}{2} {m_\chi^2}^\star \chi_L^2 + \frac{\lambda_\chi^\star}{4!} \chi_L^4+\frac{\sigma_\chi^\star}{3!} \chi_L^3 \chi_R\Bigg] \\
&\qquad + i  \Bigg[ z_{\chi\Delta}\ (\partial \chi_R).(\partial \chi_L)  + m^2_{\chi\Delta} \chi_R \chi_L+\frac{{\lambda_{\chi\Delta}}}{2!2!} \chi_R^2 \chi_L^2 \Bigg] 
\end{split}
\end{equation}
The SK interaction term corresponding to $\phi^2\chi^2$ in (\ref{twoScalarUnitaryLagrangian}) is given by,
\begin{equation}
\begin{split}
\mathcal{L}_{mix} &=
- \Bigg[\frac{h}{2!2!}\phi_R^2\chi_R^2+\frac{h_\kappa}{2!}\phi_R^2\chi_R\chi_L+\frac{h_\sigma}{2!2!}\phi_R^2\chi_L^2+\frac{h_\rho}{2!}\chi_R^2\phi_R\phi_L
\Bigg] \\
&\qquad+ \Bigg[\frac{h^\star }{2!2!}\phi_L^2\chi_L^2+\frac{h_\kappa^\star }{2!}\phi_L^2\chi_R\chi_L+\frac{h_\sigma^\star }{2!2!}\phi_L^2\chi_R^2+\frac{h_\rho^\star }{2!}\chi_L^2\phi_R\phi_L
\Bigg] \\
&\qquad+ih_\Delta \phi_R\phi_L\chi_R\chi_L
\end{split}
\end{equation}
In the above Lagrangian we have exhausted all of the interaction terms which obeys the symmetry \eqref{z2z2symmetry}, that can appear between fields on the right and left branch of the SK-contour. 

As we have already discussed in section \ref{sec:bjropenyukawa}, if we write the action in Lindblad form then the coupling constants follow certain relations (Lindblad conditions) among themselves. These conditions are the following
\begin{equation}
\begin{split}
& z_{\Delta\phi,\chi} = \text{Im}\ z_{\phi,\chi}\,,\\
& m^2_{\Delta\phi,\chi} = \text{Im}\ m_{\phi,\chi}^2\,,\\
& \text{Im}\ \lambda_{\phi,\chi} + 4 \text{Im}\ \sigma_{\phi,\chi} - 3\lambda_{\Delta\phi,\chi} = 0 \,,\\
& \text{Im } h + 2\, \text{Im } h_\kappa +\text{Im } h_\sigma + 2\, \text{Im } h_\rho = 2 h_\Delta\,.
\end{split}
\end{equation}
We have already encountered the first three conditions in section \ref{sec:bjropenyukawa}. The fourth condition encodes the relation among all of the coupling to $\phi^2\chi^2$ terms in the action.

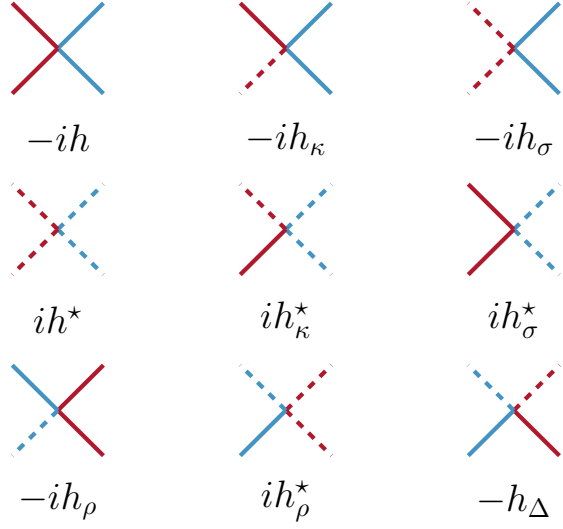
\begin{figure}[ht] 
\begin{center}
\begin{tikzpicture}[line width=1 pt, scale=0.6]
 
\begin{scope}[shift={(0,4)}]
\draw [phir, ultra thick]  (0,0) -- (1,1);
\draw [phir, ultra thick]  (0,0) -- (1,-1);
\draw [chir, ultra thick]  (0,0) -- (-1,1);
\draw [chir, ultra thick]  (0,0) -- (-1,-1);
\node at (0,-2) {\Large $-ih $};    
\end{scope} 

\begin{scope}[shift={(0,0)}]
\draw [phil, ultra thick]  (0,0) -- (1,1);
\draw [phil, ultra thick]  (0,0) -- (1,-1);
\draw [chil, ultra thick]  (0,0) -- (-1,1);
\draw [chil, ultra thick]  (0,0) -- (-1,-1);
\node at (0,-2) {\Large $ih^\star $};    
\end{scope}

\begin{scope}[shift={(5,4)}]
\draw [phir, ultra thick]  (0,0) -- (1,1);
\draw [phir, ultra thick]  (0,0) -- (1,-1);
\draw [chir, ultra thick]  (0,0) -- (-1,1);
\draw [chil, ultra thick]  (0,0) -- (-1,-1);
\node at (0,-2) {\Large $-ih_\kappa$};    
\end{scope}

\begin{scope}[shift={(5,0)}] 
\draw [phil, ultra thick]  (0,0) -- (1,1);
\draw [phil, ultra thick]  (0,0) -- (1,-1);
\draw [chil, ultra thick]  (0,0) -- (-1,1);
\draw [chir, ultra thick]  (0,0) -- (-1,-1);
\node at (0,-2) {\Large $ih_\kappa^\star $};    
\end{scope}

\begin{scope}[shift={(10,4)}]
\draw [phir, ultra thick]  (0,0) -- (1,1);
\draw [phir, ultra thick]  (0,0) -- (1,-1);
\draw [chil, ultra thick]  (0,0) -- (-1,1);
\draw [chil, ultra thick]  (0,0) -- (-1,-1);
\node at (0,-2) {\Large $-ih_{\sigma}$};    
\end{scope}

\begin{scope}[shift={(10,0)}]
\draw [phil, ultra thick]  (0,0) -- (1,1);
\draw [phil, ultra thick]  (0,0) -- (1,-1);
\draw [chir, ultra thick]  (0,0) -- (-1,1);
\draw [chir, ultra thick]  (0,0) -- (-1,-1);
\node at (0,-2) {\Large $ih_{\sigma}^\star$};    
\end{scope}

\begin{scope}[shift={(0,-4)}]
\draw [chir, ultra thick]  (0,0) -- (1,1);
\draw [chir, ultra thick]  (0,0) -- (1,-1);
\draw [phir, ultra thick]  (0,0) -- (-1,1);
\draw [phil, ultra thick]  (0,0) -- (-1,-1);
\node at (0,-2) {\Large $-ih_\rho$};    
\end{scope}

\begin{scope}[shift={(5,-4)}] 
\draw [chil, ultra thick]  (0,0) -- (1,1);
\draw [chil, ultra thick]  (0,0) -- (1,-1);
\draw [phil, ultra thick]  (0,0) -- (-1,1);
\draw [phir, ultra thick]  (0,0) -- (-1,-1);
\node at (0,-2) {\Large $ih_\rho^\star $};    
\end{scope}

\begin{scope}[shift={(10,-4)}] 
\draw [chil, ultra thick]  (0,0) -- (1,1);
\draw [chir, ultra thick]  (0,0) -- (1,-1);
\draw [phil, ultra thick]  (0,0) -- (-1,1);
\draw [phir, ultra thick]  (0,0) -- (-1,-1);
\node at (0,-2) {\Large $-h_\Delta $};    
\end{scope}

\end{tikzpicture}
\end{center}
\caption{Feynman rules for $\phi$-$\chi$ cross couplings: The red color is used for $\chi$ field and the blue colour is for the $\phi$ field.}
\label{fig:feynmanrule3} 
\end{figure}

Here we want to focus on one loop correction to the $\phi_R^2\chi_R^2$ vertex. To be more precise, we want to see whether the non-local divergences that appear in the loop integrals actually cancel or not for a physical quantity. For this purpose, we focus on $B$ type Feynman diagrams where the internal legs have different mass. Only the cross couplings enter in these diagrams and 
the Feynman rules for the cross couping and are given in the fig. \ref{fig:feynmanrule3}.
%
 
\subsection{One loop corrections to $\phi_R^2\, \chi_R^2$ vertex}
\label{subsec:phichivertexcorrection}

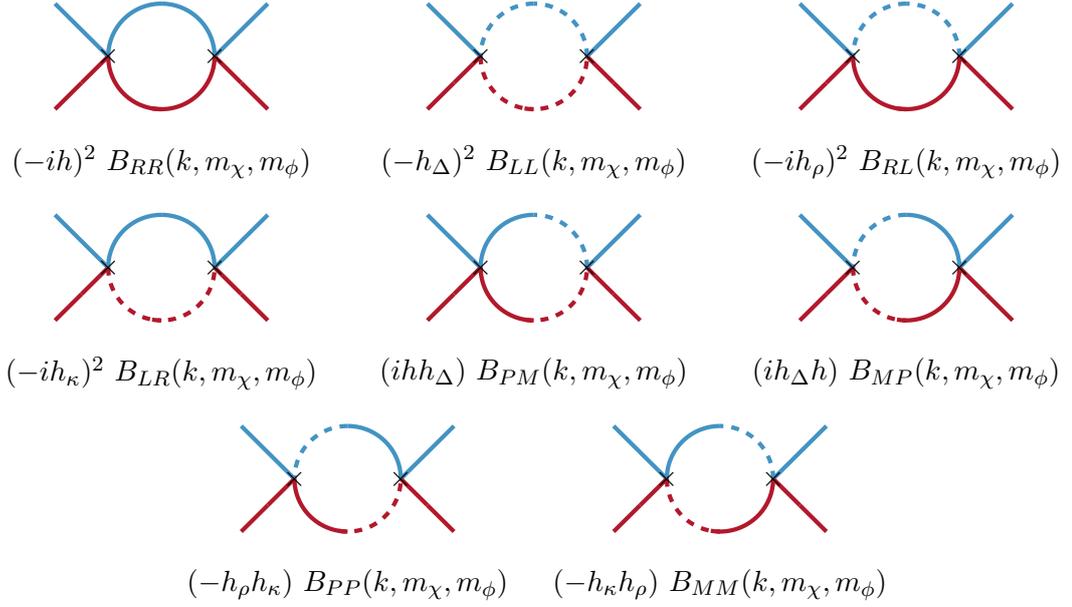
\begin{figure}
\begin{center}
\begin{tikzpicture}[line width=1 pt, scale=0.7]
 
\begin{scope}[shift={(0,0)}]
\draw [phir, ultra thick] (-2,1) -- (-1,0);
\draw [phir, ultra thick] (2,1) -- (1,0);
\chibarpropagatorr{2}{-1}{1}{0}{};
\chibarpropagatorr{-1}{0}{-2}{-1}{};
\drawphichidiagbrr{0}{0}{0}
\node at (0,-2) { $(-ih)^2\ B_{RR}(k,m_\chi,m_\phi)$};

\end{scope}

\begin{scope}[shift={(7,0)}]
\draw [phir, ultra thick] (-2,1) -- (-1,0);
\draw [phir, ultra thick] (2,1) -- (1,0);
\chibarpropagatorr{2}{-1}{1}{0}{};
\chibarpropagatorr{-1}{0}{-2}{-1}{};
\drawphichidiagbll{0}{0}{0}    
\node at (0,-2) { $(-h_\Delta)^2\  B_{LL}(k,m_\chi,m_\phi)$};
    
\end{scope}
 
\begin{scope}[shift={(14,0)}]
\draw [phir, ultra thick] (-2,1) -- (-1,0);
\draw [phir, ultra thick] (2,1) -- (1,0);
\chibarpropagatorr{2}{-1}{1}{0}{};
\chibarpropagatorr{-1}{0}{-2}{-1}{};
\drawphichidiagblr{0}{0}{0}    
\node at (0,-2) { $(-ih_\rho)^2\  B_{RL}(k,m_\chi,m_\phi)$};
    
\end{scope}

\begin{scope}[shift={(0,-4)}]
\draw [phir, ultra thick] (-2,1) -- (-1,0);
\draw [phir, ultra thick] (2,1) -- (1,0);
\chibarpropagatorr{2}{-1}{1}{0}{};
\chibarpropagatorr{-1}{0}{-2}{-1}{};
\drawphichidiagbrl{0}{0}{0}    
\node at (0,-2) { $(-ih_\kappa)^2\ B_{LR}(k,m_\chi,m_\phi)$};
    
\end{scope}

\begin{scope}[shift={(7,-4)}]
\draw [phir, ultra thick] (-2,1) -- (-1,0);
\draw [phir, ultra thick] (2,1) -- (1,0);
\chibarpropagatorr{2}{-1}{1}{0}{};
\chibarpropagatorr{-1}{0}{-2}{-1}{};
\drawphichidiagbpm{0}{0}{0}    
\node at (0,-2) { $(ihh_\Delta)\  B_{PM}(k,m_\chi,m_\phi)$};
    
\end{scope}

\begin{scope}[shift={(14,-4)}]
\draw [phir, ultra thick] (-2,1) -- (-1,0);
\draw [phir, ultra thick] (2,1) -- (1,0);
\chibarpropagatorr{2}{-1}{1}{0}{};
\chibarpropagatorr{-1}{0}{-2}{-1}{};
\drawphichidiagbmp{0}{0}{0}
\node at (0,-2) { $(ih_\Delta h)\  B_{MP}(k,m_\chi,m_\phi)$};
      
\end{scope}

\begin{scope}[shift={(3.5,-8)}]
\draw [phir, ultra thick] (-2,1) -- (-1,0);
\draw [phir, ultra thick] (2,1) -- (1,0);
\chibarpropagatorr{2}{-1}{1}{0}{};
\chibarpropagatorr{-1}{0}{-2}{-1}{};
\drawphichidiagbmm{0}{0}{0}
\node at (0,-2) { $(-h_\rho h_\kappa)\  B_{PP}(k,m_\chi,m_\phi)$};

\end{scope}

\begin{scope}[shift={(10.5,-8)}]
\draw [phir, ultra thick] (-2,1) -- (-1,0);
\draw [phir, ultra thick] (2,1) -- (1,0);
\chibarpropagatorr{2}{-1}{1}{0}{};
\chibarpropagatorr{-1}{0}{-2}{-1}{};
\drawphichidiagbpp{0}{0}{0}
\node at (0,-2) { $(-h_\kappa h_\rho)\  B_{MM}(k,m_\chi,m_\phi)$};

\end{scope}
\end{tikzpicture}
\end{center}
\caption{One loop corrections to $h$ due to $\chi$-$\phi$ loop - I.}
\label{RL_PMcontribution_to_phichi}
\end{figure}

The coupling constant corresponding to this vertices is $h$. The one loop correction to $h$ has three different kind of loop contributions - one from $\phi$ loop, one from $\chi$ loop and the rest from $\chi$-$\phi$ loop.
\begin{equation}
i\mathcal{M} = i\mathcal{M}_{\phi} + i\mathcal{M}_{\chi} + i\mathcal{M}_{\chi\phi}
\end{equation}
We have already computed $i\mathcal{M}_{\phi}$ and $i\mathcal{M}_{\chi}$ in \cite{Avinash:2017asn}; The divergences of the loop integrals are not non-local function of the external momentum thus can be cancelled by counterterms. So let us now focus on $i\mathcal{M}_{\chi\phi}$. The SK diagrams which contribute to this vertex are shown in fig. \ref{RL_PMcontribution_to_phichi} and in fig. \ref{diag:macromassrenorm3}.  

\begin{figure}
\begin{center}
\begin{tikzpicture}[line width=1 pt, scale=0.6]

\begin{scope}[shift={(0,0)}]
\draw [phir, ultra thick] (-2,1) -- (-1,0);
\draw [phir, ultra thick] (2,1) -- (1,0);
\chibarpropagatorr{2}{-1}{1}{0}{};
\chibarpropagatorr{-1}{0}{-2}{-1}{};
\drawphichidiagbrm{0}{0}{0} 
\node at (0,-2) { $(-ih)(-ih_\kappa)\ B_{PR}(k,m_\chi,m_\phi)$};

\end{scope}

\begin{scope}[shift={(10,0)}]
\draw [phir, ultra thick] (-2,1) -- (-1,0);
\draw [phir, ultra thick] (2,1) -- (1,0);
\chibarpropagatorr{2}{-1}{1}{0}{};
\chibarpropagatorr{-1}{0}{-2}{-1}{};
\drawphichidiagbmr{0}{0}{0} 
\node at (0,-2) { $(-ih_\rho)(-ih)\ B_{RP}(k,m_\chi,m_\phi)$};
    
\end{scope}

\begin{scope}[shift={(0,-4)}]
\draw [phir, ultra thick] (-2,1) -- (-1,0);
\draw [phir, ultra thick] (2,1) -- (1,0);
\chibarpropagatorr{2}{-1}{1}{0}{};
\chibarpropagatorr{-1}{0}{-2}{-1}{};
\drawphichidiagbrp{0}{0}{0}
\node at (0,-2) { $(-ih_\kappa)(-ih)\ B_{MR}(k,m_\chi,m_\phi)$};
    
\end{scope}

\begin{scope}[shift={(10,-4)}]
\draw [phir, ultra thick] (-2,1) -- (-1,0);
\draw [phir, ultra thick] (2,1) -- (1,0);
\chibarpropagatorr{2}{-1}{1}{0}{};
\chibarpropagatorr{-1}{0}{-2}{-1}{};
\drawphichidiagbpr{0}{0}{0}
\node at (0,-2) { $(-ih)(-ih_\rho)\ B_{RM}(k,m_\chi,m_\phi)$};
    
\end{scope}

\begin{scope}[shift={(0,-8)}]
\draw [phir, ultra thick] (-2,1) -- (-1,0);
\draw [phir, ultra thick] (2,1) -- (1,0);
\chibarpropagatorr{2}{-1}{1}{0}{};
\chibarpropagatorr{-1}{0}{-2}{-1}{};
\drawphichidiagblm{0}{0}{0}
\node at (0,-2) { $(-ih_\rho)(-h_\Delta)\ B_{PL}(k,m_\chi,m_\phi)$};
    
\end{scope}

\begin{scope}[shift={(10,-8)}]
\draw [phir, ultra thick] (-2,1) -- (-1,0);
\draw [phir, ultra thick] (2,1) -- (1,0);
\chibarpropagatorr{2}{-1}{1}{0}{};
\chibarpropagatorr{-1}{0}{-2}{-1}{};
\drawphichidiagbml{0}{0}{0}
\node at (0,-2) { $(-h_\Delta)(-iy_\kappa)\ B_{LP}(k,m_\chi,m_\phi)$};
    
\end{scope}

\begin{scope}[shift={(0,-12)}]
\draw [phir, ultra thick] (-2,1) -- (-1,0);
\draw [phir, ultra thick] (2,1) -- (1,0);
\chibarpropagatorr{2}{-1}{1}{0}{};
\chibarpropagatorr{-1}{0}{-2}{-1}{};
\drawphichidiagblp{0}{0}{0}
\node at (0,-2) { $(-h_\Delta)(-iy_\rho)\ B_{ML}(k,m_\chi,m_\phi)$};

\end{scope}     

\begin{scope}[shift={(10,-12)}] 
\draw [phir, ultra thick] (-2,1) -- (-1,0);
\draw [phir, ultra thick] (2,1) -- (1,0);
\chibarpropagatorr{2}{-1}{1}{0}{};
\chibarpropagatorr{-1}{0}{-2}{-1}{};
\drawphichidiagbpl{0}{0}{0}
\node at (0,-2) { $(-ih_\kappa)(-h_\Delta)\ B_{LM}(k,m_\chi,m_\phi)$};

\end{scope}

\end{tikzpicture}
\end{center}
\caption{One loop corrections to $h$ due to $\chi$-$\phi$ loop - II.}
\label{diag:macromassrenorm3}
\end{figure}
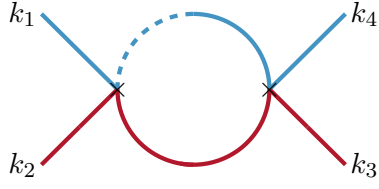
\begin{figure}
\begin{center}
\begin{tikzpicture}[line width=1 pt, scale=1.0]

\begin{scope}[shift={(10,0)}]
\draw [phir, ultra thick] (-2,1) -- (-1,0);
\draw [phir, ultra thick] (2,1) -- (1,0);
\chibarpropagatorr{2}{-1}{1}{0}{};
\chibarpropagatorr{-1}{0}{-2}{-1}{};
\drawphichidiagbmr{0}{0}{0} 
\node at (-2.25,1) {$k_1$};    
\node at (-2.25,-1) {$k_2$};
\node at (2.25,-1) {$k_3$};    
\node at (2.25,1) {$k_4$};        
\end{scope}

\end{tikzpicture}
\end{center}
 \caption{$B_{RP}(k,m_\chi,m_\phi)$, the external momenta are chosen to be ingoing.}
 \label{diag_BRP}
\end{figure}

The algebraic expression for the SK diagrams, which contribute in $\mathcal{M}_{\chi\phi}$, is given by the following equation
\begin{equation}
\label{phi2chi2}
\begin{split}
i\mathcal{M}_{\chi\phi}=& -ih \\
& + (-ih)^2 B_{RR}(k,m_\chi,m_\phi) + (-h_\Delta)^2 B_{LL}(k,m_\chi,m_\phi)\\
& + (-ih_\rho)^2 B_{RL}(k,m_\chi,m_\phi) + (-ih_\kappa)^2 B_{LR}(k,m_\chi,m_\phi)\\
& + (-ih_\rho)(-ih) B_{RP}(k,m_\chi,m_\phi) + (-ih)(-ih_\kappa) B_{PR}(k,m_\chi,m_\phi)\\
& + (-ih)(-ih_\rho) B_{RM}(k,m_\chi,m_\phi) + (-ih_\kappa)(-ih) B_{MR}(k,m_\chi,m_\phi)\\
& + (-h_\Delta)(-ih_\kappa) B_{LP}(k,m_\chi,m_\phi) + (-ih_\rho)(-h_\Delta) B_{PL}(k,m_\chi,m_\phi)\\
& +(-ih_\kappa)(-h_\Delta) B_{LM}(k,m_\chi,m_\phi) +(-h_\Delta)(-ih_\rho) B_{ML}(k,m_\chi,m_\phi)\\
& + (-ih)(-h_\Delta) B_{PM}(k,m_\chi,m_\phi) + (-h_\Delta)(-ih) B_{MP}(k,m_\chi,m_\phi) \\
& + (-ih_\rho)(-ih_\kappa) B_{PP}(k,m_\chi,m_\phi) + (-ih_\kappa)(-ih_\rho) B_{MM}(k,m_\chi,m_\phi)\\
& + (\text{one more channel})\,.
\end{split}
\end{equation}
Notice that the coupling corresponding to $B_{RP}$ is unique. So, evaluation of only $B_{RP}$ will be sufficient to show that the correction to $\phi_R^2\, \chi_R^2$ vertex has non-local divergences. The solution for this integral, in dimentional regularisation, is given in the equation below and is drawn in fig. \ref{diag_BRP}.
\begin{equation}
\begin{split}
B_{RP} &= \mu^{4-d} \int \frac{d^dp}{(2\pi)^d}\int \frac{d^dq}{(2\pi)^d} \frac{-i}{p^2+m_\chi^2-i\varepsilon} (2\pi)\delta_+(q^2+m_\phi^2) (2\pi)^d \delta^d(p-q-k)\\
&= \frac{i}{(4\pi)^2}\left(\frac{k^2-m_1^2+m_2^2}{2k^2}\frac{2}{d-4} + \text{finite terms} +\mathcal{O}(d-4)\right)\,,
\end{split}
\end{equation}
where $k=k_1+k_2$.
If the masses in the above expression are unequal then the non-local divergence can not be cancelled by a local counter-term. So if the masses of the two scalar fields  are different (and even if the tree level theory satisfies Lindblad condition), then it is not possible to remove the one loop divergences using local counter-term. We state the full correction to the $h$ vertex for completeness, which is given by the following.
\begin{equation}
\begin{split}
&(h +h_\rho + h_\kappa - ih_\Delta) (h +ih_\Delta) \Upsilon_{RR}(k_1+k_2) \\
- & (h_\rho-h_\kappa)(h +h_\rho + h_\kappa -ih_\Delta) \Upsilon_{RL}(k_1+k_2) + (k_2 \longleftrightarrow k_3)\\
+& \Big(-h\lambda_\phi + ih_\sigma^\star \lambda_{\phi\Delta} -h\sigma_\phi -h_\rho \lambda_\phi -ih_\rho \lambda_{\phi\Delta} - h_\sigma^\star \sigma_\phi \Big)\\
&\qquad \qquad\times \frac{i}{(4\pi)^2}\left[\frac{1}{d-4} + \frac{1}{2}(\gamma_E-1-\ln\,  4\pi)\right]\\
+& \Big(-\lambda_\chi h - i \lambda_{\chi\Delta} h_\sigma - \lambda_\chi h_\kappa -\sigma_\chi h + \sigma_\chi h_\sigma - i\lambda_{\chi\Delta} h_\kappa \Big)\\
&\qquad\qquad\times \frac{i}{(4\pi)^2}\left[\frac{1}{d-4} + \frac{1}{2}(\gamma_E-1-\ln\,  4\pi)\right]\,,\\
\end{split}
\end{equation}
where $\Upsilon_{RR},\,\Upsilon_{RL}$ captures the non-local divergences of all $B$-type diagrams, defined in table \eqref{Tab:divergences}. So, we find that the correction to $h$ vertex is non-local divergent.

\addcontentsline{toc}{section}{References}
\bibliographystyle{utphys} 
\bibliography{pvredjhep}   
   

\end{document}